\documentclass[preprint,journal,leqno]{vgtc}            

\onlineid{1234}

\vgtccategory{Research}

\title{Visual Decoding Operators: \\Towards a Compositional Theory of Visualization Perception}

\author{%
  \authororcid{Sheng Long}{0009-0000-9752-5898}, 
  \authororcid{Remco Chang}{0000-0002-6484-6430},
  \authororcid{Eugene Wu}{0000-0003-4254-6688}, 
  \authororcid{Alex Kale}{0000-0001-7668-2800}, and
  \authororcid{Matthew Kay}{0000-0001-9446-0419}
}

\authorfooter{
  \item
  	Sheng Long and Matthew Kay are with Northwestern University.
  	E-mail: \{shenglong, matthew.kay\}@u.northwestern.edu
  \item
  	Alex Kale is with University of Chicago. 
    Email: kalea@uchicago.edu 

  \item Eugene Wu is with Columbia University. 
  Email: ew2493@columbia.edu 
  
  \item Remco Chang is with Tufts University. 
  Email: remco@cs.tufts.edu  
}

\abstract{%
  Prior work on perceptual effectiveness has decomposed visualizations into smaller common units (e.g., channels such as angle, position, and length) to establish rankings. While useful, these decompositions lack the computational structure to predict performance for new visualization $\times$ task combinations, requiring new experiments for each. We propose an alternative unit of analysis: operationalizing quantitative visualization interpretation as sequences of composable \textit{visual decoding operators}. Using probability density function (PDF) and cumulative distribution function (CDF) charts, we examine how \chadded{four }chart-specific tasks can be decomposed into \chadded{five }reusable, chart-agnostic perceptual operations and characterize their error profiles through hierarchical Bayesian modeling. We then test generalizability by composing \chadded{one kind of }learned operators to predict performance on a structurally different task: Moritz et al.'s~\cite{moritz2023average} scatterplot mean-estimation experiment, where the chart type, chart dimensions, and analytic goal all differ from the learning conditions. With a pre-registered analysis plan, we compose operators under six candidate strategies and evaluate each against empirical data with no parameters fit to the response data. One strategy captures both bias and variance of observed responses; five alternatives fail in distinguishable ways. 
  We argue that this decoding-operator-oriented approach to empirical visualization research \chreplaced{demonstrates the feasibility of a different way of doing empirical visualization research, one where findings compose, and predictions extend beyond the conditions in which they were measured}{and theory-building lays the groundwork for generative models that can predict a distribution of likely interpretations under different viewing conditions, new chart types, and new tasks}. Free copy of this paper and supplemental materials: \href{https://osf.io/prtfq/}{\texttt{https://osf.io/prtfq}}\chdeleted{; experiment interface: \url{https://gleaming-dolphin-799fda.netlify.app/vis-decode-slider}}. 
}
  

\teaser{
  \centering
  \includegraphics[width=.9\linewidth, alt={Diagram of the approach of this paper.}]{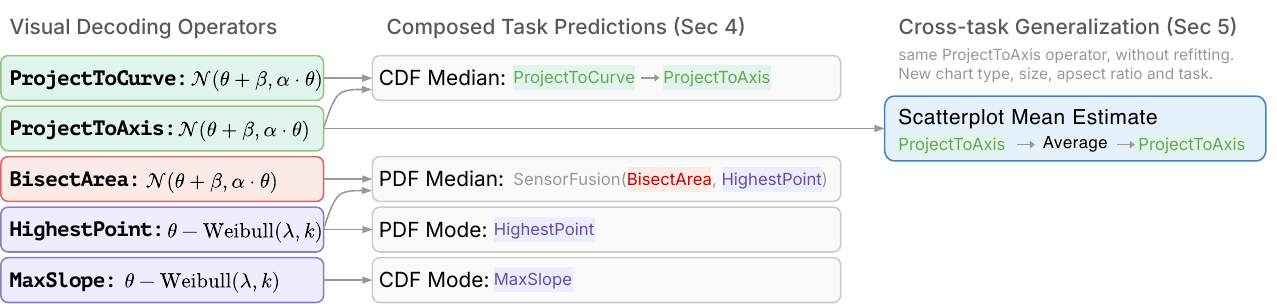}
  \caption{%
  	Visual decoding operators are perceptual primitives with estimable bias and variance that compose into task predictions (left, center). Without refitting, the same operators predict mean-estimation performance on scatterplots differing in chart type, size, and aspect ratio (right). 
  }
  \label{fig:teaser}
}

\graphicspath{{figs/}{figures/}{pictures/}{images/}{./}} 

\usepackage{tabu}                      
\usepackage{booktabs}                  
\usepackage{lipsum}                    
\usepackage{mwe}                       
\usepackage{ccicons}                   

\usepackage{amsmath}
\usepackage{newtxmath}
\usepackage{bm}
\usepackage{wrapfig}
\usepackage{pifont}
\usepackage{xcolor}
\usepackage{tcolorbox}
\newtcbox{\pill}{on line, colback=gray!15, colframe=gray!50, 
  boxrule=0.5pt, arc=4pt, boxsep=0pt, left=4pt, right=4pt, 
  top=2pt, bottom=2pt}

\usepackage{annotate-equations}
\usepackage[dvipsnames]{xcolor}
\usepackage{float}
\usepackage{soul}
\usepackage{fontawesome}

\definecolor{deepgreen}{RGB}{0, 96, 61}

\newcommand{\op}[1]{\textbf{\texttt{#1}}}

\newcommand{\opsub}[2]{\textbf{\texttt{#1}}\textsubscript{\textbf{\textsc{#2}}}}

\newcommand{\ProjToCurve}{\op{ProjectToCurve}}
\newcommand{\ProjToAxis}{\op{ProjectToAxis}}
\newcommand{\ProjToAxisX}{\opsub{ProjectToAxis}{x}}
\newcommand{\ProjToAxisY}{\opsub{ProjectToAxis}{y}}
\newcommand{\HighestPoint}{\op{HighestPoint}}
\newcommand{\BisectArea}{\op{BisectArea}}
\newcommand{\HighestPointX}{\opsub{HighestPoint}{x}}
\newcommand{\HighestPointY}{\opsub{HighestPoint}{y}}
\newcommand{\MaxSlope}{\op{MaxSlope}}
\newcommand{\Average}{\op{Average}}
\newcommand{\va}{\texttt{va}}
\newcommand{\vaX}{\texttt{va}_\textsc{X}}
\newcommand{\vaY}{\texttt{va}_\textsc{Y}}

\newcommand{\hatmuY}{{\hat\mu}_\textsc{Y}}

\usepackage{xcolor}
\usepackage[normalem]{ulem}

\newcommand{\chadded}[1]{#1}

\newcommand{\chdeleted}[1]{}

\newcommand{\chreplaced}[2]{#1}

\begin{document}



\maketitle

\section{Introduction}
Visualization research has a long tradition of decomposition: breaking the problem of understanding charts into smaller, nameable parts. Rankings order visual channels by perceptual effectiveness~\cite{cleveland1984graphical, mackinlayAutomatingDesignGraphical1986}. Task taxonomies classify what viewers accomplish with a chart~\cite{amar2005low, lee2006task}. Cognitive frameworks describe the stages of comprehension~\cite{pinkerTheoryGraphComprehension1990, kosslynUnderstandingChartsGraphs1989}. Each provides a useful organizational vocabulary, and together they have shaped design guidelines, recommendation systems, and teaching practices~\cite{moritzFormalizingVisualizationDesign2018}. Nevertheless, the \textit{units} that these decompositions produce share a common limitation: they do not capture enough internal structure to predict what will happen in situations not yet tested. This implies that every new display, task, or viewing condition requires a new experiment, an approach that does not scale.

\textit{Why is generalization so difficult?} We argue that the problem lies in the unit of analysis itself. Consider how existing decompositions carve the space. Rankings fix a task (e.g., judge a ratio) and vary the visual channel, producing an ordering over channels~\cite{cleveland1984graphical}. Task taxonomies classify what viewers accomplish with a chart (e.g., retrieve value, find extremum) but say nothing about which encoding supports each task or with what accuracy~\cite{amar2005low}. In both cases, the unit is bound to one side of the \emph{visualization $\times$ task} pairing: channels are properties of the visualization; task categories are properties of the analysis goal. Findings obtained under one pairing do not transfer to the other dimension. Knowing that position outperforms angle for ratio judgments does not tell us how either channel performs for trend estimation. Knowing that a task is classified as ``retrieve value'' does not tell us how accurately a viewer retrieves it from a bar chart versus a heatmap.

What would a more useful unit of analysis look like? It would need to be \emph{chart-agnostic}, appearing across different chart types. It would need to carry \emph{quantitative error profiles in perceptual units}, so that predictions for untested conditions can be computed rather than read off ordinal comparisons that specify direction but not magnitude. And it would need to be \emph{composable}: complex tasks expressed as sequences of operations whose error accumulates traceably, rather than evaluated as visual features where individual sources of error remain confounded.
 
In this paper, we propose such a unit of analysis: the \emph{visual decoding operator} (\cref{fig:teaser}). Operators are perceptual primitives, ranging from projecting from a point to an axis, to locating a curve's peak, to judging the steepest slope. Each operator is modeled as a function with estimable bias and variance, and a given operator (e.g., project from a known position to a curve) appears whether the chart is a Cumulative Distribution Function (CDF) or a scatterplot and whether the task is looking up a single value along a curve or estimating the mean of a cluster of points. Operators are parameterized in a perceptually uniform space (e.g.\chadded{,} visual angle) so they generalize across display contexts. And because they compose, complex tasks can be built from simpler operators, each carrying a quantified error.

Three things must hold for this approach to deliver on the generalization problem. First, operators must be defined precisely enough that their parameters (bias and variance) can be estimated in isolation, without confounding from other \chreplaced{factors}{operations} in the same task. Second, the output of one operator must serve as valid input to the next, so that error propagation through a composed sequence is tractable rather than ad hoc. Third, the resulting predictions must actually account for empirical behavior on tasks the operators were not fit to. 

Whether all three hold is an empirical question\chreplaced{.}{,} \chadded{We provide an existence proof in two stages, using structurally different visualizations and tasks, each decomposable into separable operators}\chdeleted{and we provide an existence proof based on two stages that use structurally different visualization designs and tasks, each decomposable into separable operators} (\cref{fig:teaser}). First, using Probability Density Function (PDF) and CDF charts, we design tasks that isolate individual operators and characterize their bias and variance through hierarchical modeling (\Cref{sec:build-operator}). A surprising finding is that when a task admits multiple operators (e.g., median estimation on a PDF can use both peak detection and area bisection), a \textit{sensor fusion} model that weights operators by their reliability outperforms a simple mixture model, suggesting that viewers integrate rather than select between available perceptual strategies. We then show generalizability by composing learned operators to predict performance on Moritz et al.'s~\cite{moritz2023average} scatterplot mean-estimation task (\Cref{sec:apply-operator}), a task whose \textit{chart type, dimensions, and analytic goal all differ from the conditions used to learn the operators}. \chreplaced{One o}{O}perator composition \chadded{for this task} captures observed response bias and variance \textit{with no parameters fit to the observed response data},  effectively predicting out-of-sample performance on a new visualization~$\times$~task pair based on the results of the first study. Five alternative prediction approaches fail in distinguishable ways.

\chadded{While we present here a preliminary set of five visual decoding operators and empirical predictive performance on one task, }\chdeleted{The} 
\chadded{a fully realized library of}\chdeleted{our} visual decoding operators \chadded{has the potential to} create\chdeleted{s} new opportunities for automated reasoning about visualization design. \chreplaced{More fundamentally, the empirical project of building out such a library represents a different way of accumulating knowledge in empirical visualization research, and this work serves as its initial proof of feasibility.}{For example, quantified operator-specific errors such as those we report might be used as a proxy for the cognitive effort required to decode a visualization in a new class of recommendation systems.} 
We discuss \chreplaced{what makes an operator valid, what a fuller library could enable for visualization design tools,}{implications for reshaping design patterns in tools for data interaction} as well as the new body of empirical work required to estimate \chadded{such} a library\chdeleted{ of decoding operators}.
\section{Background \& Related Work}

\subsection{Theories \& Models of Visualization Comprehension}

Rankings of visual channels~\cite{bertin1983semiology, cleveland1984graphical, kim2018assessing, harrison2014ranking} decompose the design space into an ordered list of encodings, providing easy-to-access heuristics for visualization recommendation. Task taxonomies such as Amar et al.~\cite{amar2005low} decompose analytic activity into categories (e.g., Retrieve Value, Find Extremum, Correlate) that describe \textit{what} a viewer can accomplish with a chart. Cognitive frameworks decompose the comprehension process itself into named stages~\cite{simkin1987information,pinkerTheoryGraphComprehension1990,hegarty2011cognitive, padilla2018decision}. Each of these traditions contributes a useful organizational vocabulary for describing visualization comprehension.

However, the \textit{units} these decompositions produce share a common property: they are \textit{categorical or ordinal}. Rankings reduce experiment outcomes to a single scalar, collapsing over variance in responses and masking differences across individuals and contexts~\cite{davisRisksRankingRevisiting2022}. Task taxonomies classify what a viewer does but say nothing about how accurately or with what bias they do it. Cognitive stage models name the processes involved but remain specified at a conceptual level, without operationalizing them into units whose parameters can be estimated from data.

A separate line of work has pursued more operational forms of decomposition. Gillan and Lewis's Mixed-Arithmetic Perceptual (MA-P) model~\cite{gillan1994componential} predicts that graph comprehension time scales linearly in the number of cognitive components used. Peebles and Cheng \cite{peeblesExtendingTaskAnalytic2002, peebles2003modeling} implemented process models of graph reading in the ACT-R/PM cognitive architecture, simulating the sequence of perceptual, cognitive, and motor operations a viewer executes to predict both response latency and eye fixation patterns.  These computational models produce testable quantitative predictions, but their target is response time, not the perceptual output itself (e.g., the estimate a viewer produces and the errors that arise from the visual encoding). Wu and Chang~\cite{wu2024design} decompose the visualization design process into functional transformations and visual encodings, showing that design tasks can be described as compositions of operations. Nevertheless, their decomposition applies to the encoding stage, not the subsequent interpretation by a human viewer. More recently, Huang and Nobre~\cite{huang2025vistruct} use large language models to decompose visualization tasks into sequences of analytic subtasks drawn from established task taxonomies~\cite{amar2005low}, validated against expert reasoning patterns. Their subtask units, such as ``Retrieve Value'' or ``Filter'', describe what the viewer needs to accomplish at each step, but do not characterize the perceptual operations that execute each step or the error signatures those operations produce. 

A few empirical studies have moved toward quantitative models of specific perceptual operations: Talbot et al.~\cite{talbot2012empirical} developed an empirical model of slope ratio comparison that captures aspect-ratio-dependent bias, characterizing a specific visual judgment with estimable parameters rather than an ordinal ranking. What remains missing is a framework that treats such characterized operations as composable units whose error propagates traceably through multi-step tasks. 

\subsection{Error Attribution in Visualization Experiments} 

Observed error in a visualization experiment is rarely attributable to a single source. A participant's response reflects not only their perceptual judgment but also the elicitation method used to collect it, the study design that framed the task, and individual differences in visualization literacy, domain knowledge, and task interpretation. Each of these stages can introduce error, and without a way to separate these sources, the perceptual signal of interest remains confounded with measurement artifacts.

The elicitation method is one well-documented source of error. Since Heer and Bostock~\cite{heerCrowdsourcingGraphicalPerception2010} demonstrated that graphical perception experiments could be successfully conducted online, crowdsourced studies have become the norm, and researchers have developed a range of \textit{interactive response formats}, from clicking directly on the perceived centroid of the stimuli~\cite{hongWeightedAverageIllusion2021} to dragging response probes~\cite{xiongBiasedAveragePosition2020a}. The choice of elicitation format is not neutral: numeric entry produces rounding behaviors~\cite{cleveland1984graphical, hondaNumberBiasWisdom2022}, and sketching may be confounded with drawing ability and task interpretation alongside perceptual judgments~\cite{promaVisualStenographyFeature2025,nobre2024reading}. 

Other sources of error operate upstream of the response. Nobre et al.~\cite{nobre2024reading} found that participants who misread visualizations failed at distinct cognitive stages, such as misreading channels, confusing labels with values, or failing to translate questions into visual queries. Sarma et al.~\cite{sarma2024tasks} argue complementarily that aspects of study design frequently treated as incidental, such as onboarding procedures, tutorials, and training phases, can interact with individual-level factors such as visualization literacy in ways that are difficult to separate from the perceptual signal itself.

To build componential models that predict the perceptual output of individual visual operations, we need an approach that both decomposes tasks into separable operators and explicitly controls for non-perceptual error sources. \Cref{sec:the_model} introduces the decomposition framework, and \Cref{sec:methodology} details how our experiment design addresses the sources of error identified above.

\section{A Model For Visual Operators}
\label{sec:the_model}

\begin{figure}[!htbp]
    \centering
    \includegraphics[width=0.9\linewidth]{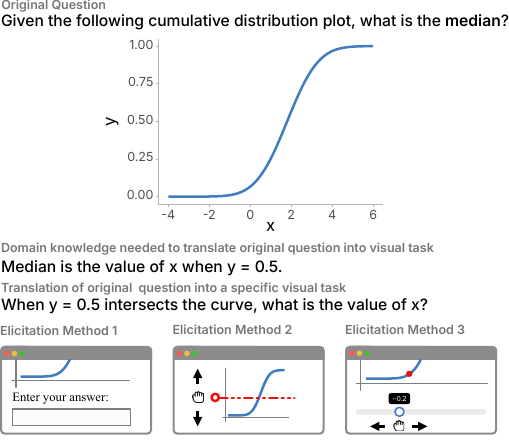}
    \caption{An illustrative example of different ways tasks can be set up to elicit a response. We use elicitation method 3 (details in \Cref{sec:methodology}).
    }
    \label{fig:modeling-approach}
\end{figure}

Consider this scenario: you ask a participant to identify the median of a cumulative distribution function (CDF; $y = \Pr(X \leq x)$). A typical visualization study might elicit estimates of this value by showing participants different CDFs and asking them to report the medians using different elicitation methods (\Cref{fig:modeling-approach}). Researchers then calculate the error and variance in participants' ability to identify medians from CDFs. As a starting point, assume the participant's response is drawn from some distribution: 
\begin{align}
\label{eq:model_eq}
    \text{response} \sim \mathcal{N}(\text{true value} + \text{bias}, \text{standard deviation})
\end{align}

\begin{figure}[H]
    \centering
    \includegraphics[width=.9\linewidth]{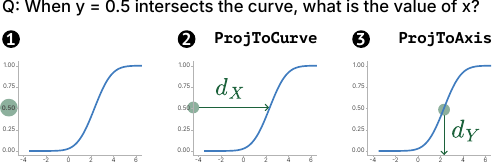}
    \caption{The task decomposition process using median identification on CDF as a concrete example.}
    \label{fig:model_diagram}
\end{figure}

But this treats the task as monolithic --- it does not distinguish where error enters. Instead, our task instructions (\Cref{sec:method_task}) decompose CDF median identification into three distinct steps that follow from \chadded{a correct interpretation of} the task \chdeleted{definition}itself (\Cref{fig:model_diagram}): \ding{202} locate $y = 0.5$ on the Y-Axis, \ding{203} perform a horizontal visual scan, projecting horizontally from $y = 0.5$ to where it intersects the CDF curve (\ProjToCurve), then \ding{204} project vertically from the curve intersection to the X-Axis (\ProjToAxisX). Given that the label ``0.5'' is aligned and next to its position on the y axis, \ding{202} is trivial. The core challenge then becomes measuring the error in the horizontal projection (\ding{203}) and vertical projection (\ding{204}). 

Before we can specify these operators formally, we need to establish \textit{what space we measure error in}. 
To create a model that generalizes across viewing context, from \chreplaced{laptops to desktop monitors}{desktop screens to museum installations}, we need to measure error in a standardized, \textit{viewing-context-independent} space. We choose to measure error in \textit{visual angle space}, which is the angle a distance subtends at the eye. Visual angles are typically used in vision science research~\cite{mather2016foundations} and allow us to approximate the perception of physical distances
regardless of viewing conditions\chreplaced{:}{. Values are transformed from data space to physical space, and then to visual angle space (measured in degrees) through}
\begin{align}
\label{eq:vis_angle}
    \text{visual angle of object} = 2 \times \tan^{-1}\left(\frac{\text{physical size of object}}{2 \times \text{viewing distance}} \right)
\end{align} 

To calculate the visual angle of on-screen stimuli, which requires knowing the viewing distance (\Cref{eq:vis_angle}), we implemented the ``virtual chinrest'' method proposed by Li et al. \cite{liControllingParticipantsViewing2020}. This browser-based method measures a participant’s viewing distance by two calibration tasks: 1) a \textit{blind spot} task, where participants find their physiological blind spot to establish their distance from the screen, and 2) an \textit{object scaling} task, where they match a digital rectangle to a physical object of known size (like a credit card) to find the monitor's pixels-per-centimeter ratio. We reimplemented these tasks on the reVISit platform~\cite{cutler2026revisit,dingReVISitSupportingScalable2023}, following its open-source version.\footnote{\url{https://github.com/QishengLi/virtual_chinrest}} In the paper we will denote visual angle transformations as:
\begin{align}
\va(d; \mathit{context})
\end{align}
Where $d$ is a value in data space and $\mathit{context}$ includes all participant- and context-specific parameters needed to transform a value from data space to visual angle, including distance to screen, pixels-per-centimeter, and data units-per-pixel.\footnote{For simplicity we typically omit $\mathit{context}$ and write $\vaX(\cdot)$ or $\vaY(\cdot)$ to refer to the context-specific visual angle transformation along the $x$ or $y$ axes.}

\begin{figure}[!htbp]
    \centering
    \includegraphics[width=0.95\linewidth]{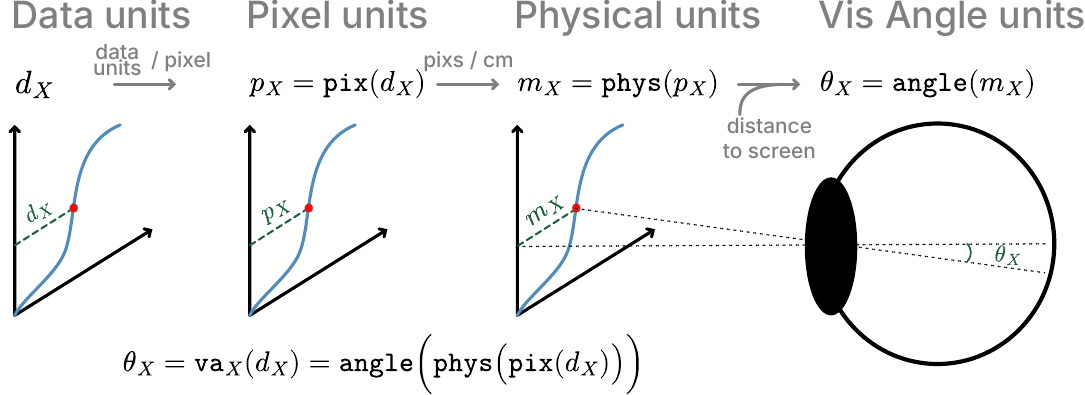}
    \caption{\chadded{How values in data units are translated to visual angle units, making the measurement invariant to individual viewing conditions and axis relabelings that leave the rendered geometry unchanged.}}
    \label{fig:vis_angle_illustration}
\end{figure}

\begin{figure}[!htbp]
    \centering
    \includegraphics[width=0.9\linewidth]{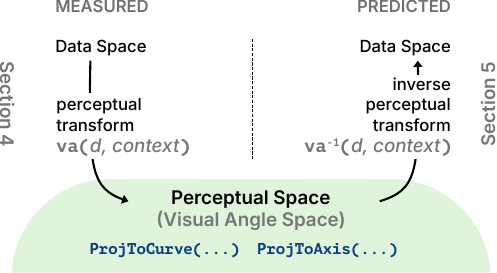}
    \caption{A horseshoe diagram~\cite{varona2025theory} of this paper's approach to modeling.}
    \label{fig:modeling-approach-overall}
\end{figure}

With visual angle space defined, we denote the true value in this space as $\theta$\chadded{, projection distance in this space as $\Delta_\theta$} and rewrite the two core operators (\ding{203}, \ding{204}) in the median-finding decomposition as: 
\vspace{0.5em}
\begin{align*}
\\
    \tag{\ding{203}} \ProjToCurve(\eqnmarkbox[deepgreen]{node3}{\theta}, \eqnmarkbox[deepgreen]{node6}{\Delta_\theta}) & \; \sim \;\; \mathcal{N}(\eqnmarkbox[deepgreen]{node4}{\theta} + \eqnmarkbox[deepgreen]{node1}{\beta_{\texttt{curve}}}, \eqnmarkbox[deepgreen]{node2}{\sigma_{\texttt{curve}}}) \\ \\ \\ \\ 
\tag{\ding{204}} \ProjToAxisX(\theta_X, \theta_Y) &\sim \mathcal{N}(\theta_X + \beta_{\texttt{axis}}, \sigma_{\texttt{axis}})
\end{align*}
\annotate[yshift=-1em]{below,left}{node1}{\textbf{\small bias for} \ProjToCurve} 
\annotate[yshift=-2.5em]{below,left}{node2}{\textbf{\small standard deviation for} \ProjToCurve}
\annotatetwo[yshift=0.5em]{above}{node3}{node4}{\textbf{\small true value in visual angle space}}
\annotate[yshift=0.1em,xshift=-0.3em]{below,left}{node6}{\textbf{\small projection distance for} $\theta$}
\noindent Decomposing the task into visual operators becomes:
\begin{align*}
    & \text{user's observed response} = \vaY^{-1} \\ \\  
    & \left[{\small \ProjToAxisX}\left({\small \ProjToCurve}
    \left(\eqnmarkbox[deepgreen]{node1}{\vaY(0.5)}, \eqnmarkbox[deepgreen]{node2}{\vaX(d_\textsc{x})}\right), \eqnmarkbox[deepgreen]{node3}{\vaY(d_\textsc{y})}\right)\right]
\end{align*}
\annotate[yshift=0em]{below,left}{node1}{\textbf{\small visual angle of} $y = 0.5$} 
\annotate[yshift=0.5em]{left}{node2}{\textbf{\small visual angle of distance to curve}} 
\annotate[yshift=-1.5em]{below, left}{node3}{\textbf{\small visual angle of distance to axis}}
\vspace{1.5em}

\noindent ... where $\vaY^{-1}(\cdot)$ is the inverse visual angle function that transforms a value in visual angle space back to data values along the $y$ axis, given relevant context-specific parameters \chadded{(see also \Cref{fig:modeling-approach-overall})}. 

Ultimately, we aim to demonstrate that under this modeling approach, we can decompose tasks into and generate estimates of individual decoding operators (\Cref{sec:build-operator}), combine different operators (\Cref{sec:Task1}), and finally generate predictions for new experimental conditions and questions that the existing set of operators are not trained on (\Cref{sec:apply-operator}).

\section{Building Operators}
\label{sec:build-operator}

In the following section, we detail our experimental approach for addressing two methodological challenges: 1) ensuring participants perform the intended visual operator(s) to minimize the effects of the identifiability problem and 2) obtaining measurements in online crowd-sourced studies, such as the physical distance to the screen and participants' responses as outcomes of visual decoding operators. 

\subsection{Experiment Design}
\label{sec:methodology}

\subsubsection{Tasks, Instructions, \& Training}
\label{sec:method_task}

We use both probability density function (PDF) charts and cumulative distribution function (CDF) charts, focusing on \textbf{median} and \textbf{mode} identifications. 
We selected these statistics because the process of reading median and mode can be decomposed into separable visual operators, as demonstrated in \Cref{sec:the_model}. 

A challenge in prior work~\cite{ibrekk1987graphical} is that tasks often require participants to already understand statistical concepts. Errors in such tasks may reflect misunderstanding of definitions rather than imprecision in perception. Thus, we avoided directly asking for statistical measures and instead decomposed abstract tasks into concrete graphical operations (\Cref{tab:exp_instruct}). More concretely, rather than asking for ``the median,'' our instructions \textbf{specify the exact graphical components and actions required (effectively, the visual operators to use)}. For example, when eliciting median estimates from a CDF, we instruct participants to position a red dot where $y = 0.5$ intersects the CDF curve.

\begin{table}[!htbp]
    \centering
    \begin{tabular}{p{3.5cm} p{4.5cm}}
    \toprule
     \chreplaced{\textbf{Operator \& Example Task}}{\textbf{Task}} & \textbf{Experiment Instruction}  \\ \midrule 
     \chreplaced{\HighestPoint}{PDF Mode} \newline \cref{sec:task2} \chreplaced{(PDF Mode)}{\HighestPoint}  & \textit{``Drag the slider to move the ball along the curve to the highest point (where y is greatest).''} \\ 
     \BisectArea \newline 
     \cref{sec:Task1} (PDF Median) & \textit{``Drag the slider to move the ball along the curve until a hypothetical vertical line through it divides the area into two equal halves.''} \\ 
     \MaxSlope \newline 
     \cref{sec:slope-judgment} (CDF Mode)  & \textit{``Drag the slider to move the ball along the curve to where the slope is steepest.''} \\ 
     \ProjToCurve \newline 
     \cref{sec:projection-task} (CDF Median) & \textit{``Drag the slider to move the ball along the curve until it rests where the curve crosses y=0.5.''} \\ \ProjToAxisX, \ProjToAxisY\ & \textit{``Drag the sliders such that the two balls on the x-axis and y-axis match the red dot.''}\\  
     \bottomrule
    \end{tabular}
    \caption{\chreplaced{Operators, t}{T}asks and their respective instructions used during experiment.}
    \label{tab:exp_instruct}
\end{table}

To further minimize confounding factors such as instruction misunderstanding, we provided screen recordings of intended behavior and training trials for each task. This approach allows us to attribute errors to the visual operators themselves rather than to task comprehension. Given our precise characterization of the error due to visual operators, future experiments should be able to precisely measure the additional error due to participants incorrectly translating the task to those operators (due to lack of prior knowledge or visualization literacy).

\subsubsection{Stimuli \& Apparatus}
\label{sec:method_stimuli}

We used the skewed generalized $t$ distribution~\cite{theodossiouFinancialDataSkewed1998} as our data-generating distribution for its ability to independently control skewness, kurtosis, and scale (details in~\Cref{sec:skew-t-pdf}). Parameters of the SGT distribution were sampled such that the target statistics (i.e., median and mode) vary in the ranges of $x \in [-5, 5], y \in [0, 1]$. Stimuli data were generated using \texttt{stdlib-js}~\cite{stdlib} and rendered with \texttt{D3} ~\cite{bostockD3DataDrivenDocuments2011} at dimensions of 600 px $\times$ 450 px (details in \Cref{sec:experiment-interface}). 

To minimize potential confounds, we implemented a custom elicitation interface \chadded{--- }\chdeleted{with several key design features. We used }a slider with no default starting position to prevent anchoring effects\chadded{ (see \Cref{sec:slider-interface} for details)}.\chdeleted{ The slider was positioned away from the chart axes to prevent participants from directly aligning the slider position with axis values, encouraging them to focus on the red dot annotation that appears on the curve itself. Participants first clicked anywhere on the slider to initialize it (\textsf{a}), which displayed a red dot annotation on the curve at the corresponding position (\textsf{b}). They could then control the red dot's position by dragging the slider or using left and right arrow keys (\textsf{c}).} We chose this slider-based approach over alternatives, such as direct clicking on the curve, because it eliminated ambiguity about the source of error while constraining participants to perform the specific visual operators under investigation. 
We used the ReVISit platform~\cite{dingReVISitSupportingScalable2023,cutler2026revisit} to elicit responses from participants (\href{https://mika-long.github.io/vis-decode/vis-decode-slider/}{\faExternalLinkSquare\ example interface}). 

\subsubsection{Experiment Procedure and Participants}
\label{sec:exp1-procedure}
\begin{figure}[!htbp]
    \centering
    \includegraphics[width=0.95\linewidth]{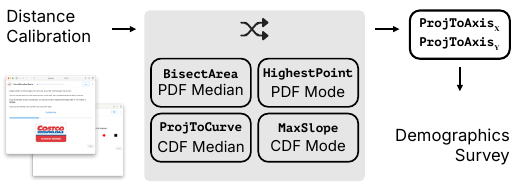}
    \caption{Experiment procedure. The study used a within-subjects design.}
    \label{fig:process}
\end{figure}
After consenting on the experiment website, participants first performed \textit{calibration tasks} to measure screen-to-pixel ratio and eye-to-screen distance. The experiment consisted of five blocks, one for each task. For each task, participants first viewed instructional screens and a short video clip explaining the task, completed two training trials with fixed stimuli across participants, and then performed ten testing trials with randomized stimuli, resulting in a total of $5 \times (10 + 2) = 60$ trials per participant. Finally, participants provided demographic information and optional text feedback on their strategies and the clarity of instructions.

We recruited 24 participants in an IRB-approved study \chadded{(STU00223203)}\chdeleted{[removed for anonymization]} on \texttt{Prolific.co} who are between the ages of 18 and 65, have normal or corrected-to-normal vision, speak fluent English, and reside in the United States. \chadded{We excluded 8 participants who have measured eye-to-screen distance less than 20 cm or have Pearson correlation < 0.5 to the answers, as typical participants showed correlations $> 0.95$ and maintained the expected viewing distance of approximately $50$ cm.\footnote{This exclusion criteria was not pre-registered; we applied the same exclusion criteria to the experiment we ran in~\Cref{sec:apply-operator}, where they were pre-registered.}} The study was distributed to a \chadded{gender-}balanced sample of participants. \chadded{Participants ranged in age from 31 to 63 (M = 43.6, SD = 9.7).} We paid each participant \$4.86 for completing the experiment, with a potential bonus of up to \$1.50 depending on the participants' precision. The experiment took on average 18.6 minutes to finish, resulting in an hourly wage of \$20.51 / hr\chdeleted{, exceeding the US federal minimum wage}.

\subsubsection{Analysis Approach}
\label{sec:exp1-analysis}
We use multi-level Bayesian regression to analyze our results. Specifically, we fit models using values transformed from physical space (cm) into visual angle space, unless otherwise specified. We excluded participants with Pearson correlation $< 0.5$ to the actual answer or have eye-to-screen distance $< 20$ cm, as typical participants showed correlations $> 0.95$ and maintained the expected viewing distance of approximately $50$ cm. We fit five Bayesian hierarchical mixed effects models, with all parameters achieving bulk effective sample size (\texttt{Bulk\_ESS}) and tail effective sample size (\texttt{Tail\_ESS}) values ranging from $1976$ to $19940$. We assess convergence using the Gelman-Rubin diagnostic ($\hat{R} =1.00$ for all population-level parameters). We assess all model fits using posterior predictive checks following the recommendations of S{\"a}ilynoja et al.~\cite{sailynoja2025recommendations}. We calculate and present our results with \texttt{R} packages \texttt{brms}~\cite{brms}, \texttt{tidybayes}~\cite{tidybayes}, \texttt{ggdist}~\cite{kay2023ggdist}, \texttt{bayesplot}~\cite{bayesplot}, \texttt{sgt}~\cite{sgt}, \texttt{cmdstanr}~\cite{cmdstanr}, \texttt{ggplot2}~\cite{ggplot2}, \texttt{modelr}~\cite{modelr}, and \texttt{tidyverse}~\cite{tidyverse}. 

\subsection{Results}
\label{sec:operator_results}

We characterize five operators across four tasks: projection operators (\ProjToCurve, \ProjToAxisX, \ProjToAxisY) from CDF median identification (\Cref{sec:projection-task}), \HighestPoint\ from PDF mode (\Cref{sec:task2}), \MaxSlope\ from CDF mode (\Cref{sec:slope-judgment}), and \BisectArea\ from PDF median (\Cref{sec:Task1}).

\subsubsection{Projection Operators}
\label{sec:projection-task}

\ProjToCurve\ then \ProjToAxisX, i.e., median identification on a CDF  (\Cref{fig:model_diagram}), involves projecting from $y = 0.5$ to its intersection on the CDF curve and then projecting from that intersection to the X-Axis. We model all projection-related operators as Normal distributions over values parameterized in visual angle on the respective axis. For all projection tasks, we hypothesize that the variance of the operator's output increases with the visual angle distance between the projection start point and end point.

The following model is for \ProjToAxisX\ for a particular participant $j$:
\begin{align*}
    \ProjToAxisX(\eqnmarkbox[deepgreen]{t1}{\theta_X}, \theta_Y) &\sim \mathcal{N}\left(\eqnmarkbox[deepgreen]{t2}{\theta_X} + \beta_{j}, \sigma_{j}\right)   
\end{align*}
\annotatetwo[yshift=0.8em]{above}{t1}{t2}{\textbf{\small visual angle of true value}}
\noindent The \ProjToAxisX\ operator returns a Normal distribution centered at the true value plus a participant-specific bias $\beta_j$, capturing whether participant $j$ systematically over- or under-estimates the projected position. The observation-specific standard deviation $\sigma_j$ captures how much the participant's responses spread for a given trial:
\begin{align*}
    \sigma_{j} &= \alpha_j \cdot \theta_Y 
\end{align*}
\noindent  The standard deviation scales multiplicatively with how far the point must be projected, i.e.\chadded{,} the visual angle distance to $x$-axis, $\theta_Y$, where $\alpha_j$ acts as a scaling parameter for participant $j$. We compared additive and multiplicative scaling relationships on pilot data; the multiplicative specification yielded better posterior predictive calibration and is used throughout. The models for \ProjToCurve\ and \ProjToAxisY\ are defined similarly as the model above. \chadded{Note that these models condition variance on projection distance but not on the local slope at the projection target; incorporating slope-dependent terms via standard first-order error propagation is a natural future refinement.} Details of the full hierarchical specifications (which model population means and standard deviations of $\beta_j$ and $\alpha_j$) are in~\Cref{sec:supplemental_materials}.

\begin{figure}[!htbp]
    \centering
    \includegraphics[width = \linewidth]{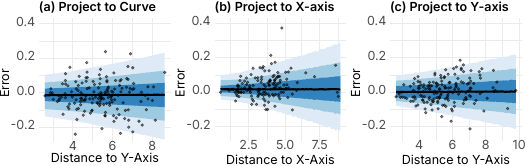}
    \caption{Plots for (a) \ProjToCurve, (b) \ProjToAxisX, and (c) \ProjToAxisY: error vs. distance between start and end points of projection, where points show observed errors with blue bands representing 50\%, 80\%, and 95\% posterior predictive intervals, marginalizing over participants.}
    \label{fig:projection-overall}
\end{figure}

\Cref{fig:projection-overall} shows observed errors plotted against projection distance for all three projection operators, with posterior predictive intervals marginalizing over participants. In all three cases, the spread of errors increases with distance, consistent with the multiplicative scaling specification. The posterior predictive intervals cover the observed data well, with most points falling within the 95\% interval. 

\begin{figure}[!htbp]
    \centering
    \begin{subfigure}[b]{0.6\linewidth}
        \centering
    \includegraphics[width=\linewidth]{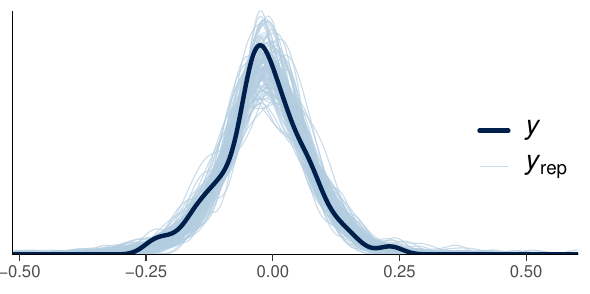}
    \caption{}
    \label{fig:task3-pp-check}
    \end{subfigure}
    \hfill
    \begin{subfigure}[b]{0.35\linewidth}
        \centering
    \includegraphics[width=\linewidth]{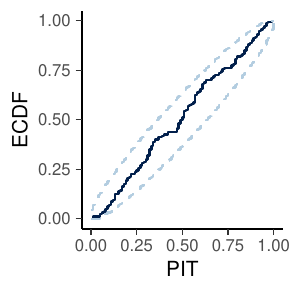}
    \caption{}
    \label{fig:task3-pit-ecdf}
    \end{subfigure}
    \caption{Posterior predictive checks for \textbf{\texttt{ProjectToCurve}}. (a) KDE density overlay plot with 100 posterior draws. (b) PIT-ECDF Calibration plot. }
    \label{fig:projtocurve-ppcheck}
\end{figure}

We assess model fit using two standard posterior predictive checks (\Cref{fig:projtocurve-ppcheck}). In the density overlay (\Cref{fig:projtocurve-ppcheck} (a)), the dark line shows the observed distribution of responses and each lighter line is a density drawn from the model's posterior predictive distribution. If the model is adequate, the observed density should fall among the predicted densities in location, spread, and shape. The PIT-ECDF plot (\Cref{fig:projtocurve-ppcheck} (b)) transforms each observation by its posterior predictive CDF and plots the resulting empirical CDF against a uniform reference. Adequate calibration means the empirical CDF falls within the simultaneous confidence envelope. For \ProjToCurve, both diagnostics indicate adequate fit. Analogous checks for the remaining operators appear in \Cref{sec:additional-posterior-checks}.

\subsubsection{Highest Point Operator}
\label{sec:task2}

\begin{wrapfigure}[9]{l}{0.33\linewidth}
  \vspace{-2em}
  \begin{center}
    \includegraphics[width=0.17\textwidth]{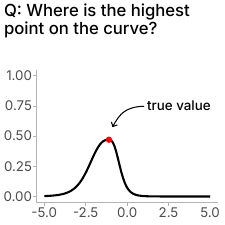}
  \end{center}
\end{wrapfigure}

\HighestPoint, i.e., mode identification on a PDF, asks participants to identify the highest point on the curve. While participants generally demonstrated high precision with this task, the distribution of errors exhibits a heavy tail (\Cref{sec:pp-highestpointY}). To capture this pattern effectively, we model the operator as returning the true value in visual angle $\theta$ minus a non-negative error term $\epsilon$: 
\begin{align*}
    \HighestPointY(\theta) &= \theta - \epsilon
\end{align*}
The error term follows a Weibull participant-specific distribution: 
\begin{align*}
    \epsilon &\sim \text{Weibull}(\lambda_{j}, k_{j})
\end{align*}
where $\lambda_j > 0$ and $k_j > 0$ are the scale and shape parameters for participant $j$ respectively. The Weibull distribution generalizes the Exponential distribution through an additional shape parameter, allowing the rate of tail decay to vary. We selected it over other heavy-tailed candidates based on leave-one-out cross-validation (\Cref{sec:supplemental_materials}). 

\begin{figure}[!htbp]
    \centering
    \includegraphics[width=0.8\linewidth]{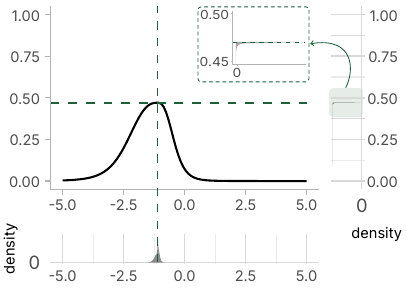}
    \caption{Posterior predictive error distributions for \HighestPointY\ applied to an example stimulus. The top-left panel shows the stimulus curve with the true peak marked. The bottom-left panel shows the marginal predictive distribution of error along the x-axis, and the top-right panel shows the corresponding distribution along the y-axis (\HighestPointY). }
    \label{fig:task2-overall-y}
\end{figure}

When the curve is relatively flat near the peak, even small errors along the $y$-axis can correspond to large displacements along the $x$-axis (\Cref{fig:task2-overall-y}). We derive a model for $x$-axis error from the $y$-axis error through the geometry of the stimulus curve. In \Cref{sec:Task1} we provide an alternative model for x-axis error for \HighestPoint\ using a Normal distribution for tractability within the sensor fusion framework. 

\subsubsection{Max Slope Operator}
\label{sec:slope-judgment}

\begin{wrapfigure}[9]{l}{0.33\linewidth}
  \vspace{-2em}
  \begin{center}
    \includegraphics[width=0.18\textwidth]{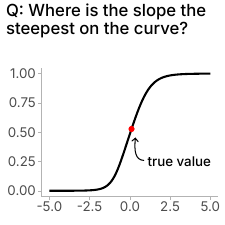}
  \end{center}
\end{wrapfigure}
\MaxSlope, i.e., mode identification on a CDF, requires judging the point on the curve whose tangent has the maximum slope. We express slope in visual angle space by deriving the mapping between $\Delta x / \Delta y$ in data space and visual angle space. Similar to \HighestPointY, we model the \MaxSlope\ operator as returning the true value minus a non-negative error term $\epsilon$: 
\begin{align*}
    \MaxSlope(\theta) &= \theta - \epsilon
\end{align*}

Because the true answer is the maximum slope on the curve, any other response necessarily has a smaller value. Participants thus tend to underestimate the maximum slope, and the distribution of errors is asymmetric with a heavy tail. We thus model this error $\epsilon$ using a Weibull distribution, selected over other heavy-tailed candidates via leave-one-out cross-validation (see supplementary material):
\begin{align*}
    \epsilon &\sim \text{Weibull}(\lambda_{j}, k_{j})
\end{align*}
where $\lambda_j > 0$ and $k_j > 0$ are the scale and shape parameters for participant $j$ respectively. 

\begin{figure}[H]
    \centering
    \includegraphics[width=0.9\linewidth]{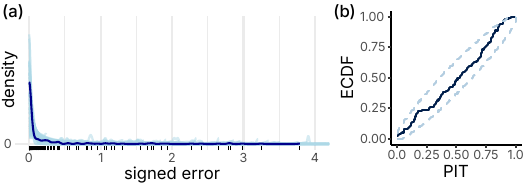}
    \caption{(a) Posterior retrodictive density of participants' errors overlaid with density of observed errors, plotted on top of strip plot of observed errors for \MaxSlope. (b) PIT-ECDF Calibration plot. }
    \label{fig:task3-ppcheck}
\end{figure}

\subsubsection{Bisect Area Operator}
\label{sec:Task1}

\begin{wrapfigure}[10]{l}{0.33\linewidth}
  \vspace{-2em}
  \begin{center}
    \includegraphics[width=0.18\textwidth]{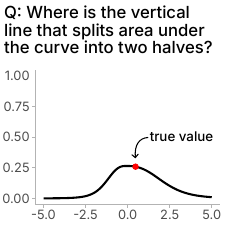}
  \end{center}
\end{wrapfigure} 

In pilots, we observed non-uniform errors, and posited that due to the difficulty of the task, participants may use two different operators to accomplish this task. For unimodal distributions like those in our experiment, when the distribution is \textit{symmetric}, the mode and median are equal and one can use the low-error \HighestPointX\ operator. Even if skewness is low, the mode may still be a good approximation of the median considering the low perceptual error of \HighestPointX. As described in \Cref{sec:task2}, $x$-axis error is derived from $y$-axis error through the stimulus curve's geometry and approximated as Normal:
\begin{align*}
    \HighestPointX(\theta) &\sim \mathcal{N}(\theta + \beta^\texttt{HP}_{j}, \sigma^\texttt{HP}_{j}) 
\end{align*}
When skewness is high, one may have to resort to a higher-error \BisectArea\ operator, visually estimating the point that divides the region into two equal-area halves.
\begin{align*}
    \BisectArea(\theta)  &\sim \mathcal{N}(\theta + \beta^\texttt{BA}_{j}, \sigma^\texttt{BA}_{j}) 
\end{align*}
But how \textit{precisely} do people select when to use which operator? One natural hypothesis is that participants probabilistically select between the two operators on each trial (a \textit{mixture} model). However, posterior predictive checks of a mixture model do not adequately capture the observed response distribution (see supplementary material).

Alternatively, prior work~\cite{simkin1987information} and our own observation suggests that an \textit{anchoring-and-adjustment} process could be at work: perhaps participants first identify the mode of the PDF (anchoring), then adjust based on the skewness of the distribution. We model this by combining estimates from both operators using \textit{inverse mean squared error weighting}, a technique commonly used in sensor fusion~\cite{elmenreich2002introduction} for giving more weight to more reliable sensors~\cite{wang2017study}. Specifically, we assume that for the combined operator, which we dub \textbf{\texttt{BAHP}}, participants' selections follow a weighted mean of the other two operators:
\begin{align*}
    \smash{\underbrace{\textbf{\texttt{BAHP}}}_{\equiv\small \texttt{BisectAreaHighestPoint}}}(\theta_\texttt{median}, \theta_\texttt{mode}) = 
    \eqnmarkbox[deepgreen]{p1}{w} \cdot  \BisectArea(\theta_\texttt{median}) \\ + (1 - \eqnmarkbox[deepgreen]{p2}{w}) \cdot \HighestPointX(\theta_\texttt{mode}) \\ 
\end{align*}
\annotatetwo[xshift=-3em, yshift=-1.8em,]{below, label below}{p1}{p2}{\textbf{\small how much a person weights one operator versus another in their estimate}}
\noindent \BisectArea\ targets the median directly, so its mean squared error (MSE) reflects only its own bias and variance:
\begin{align*}
\text{MSE}(\BisectArea) = (\beta^\texttt{BA}_j)^2 + (\sigma^\texttt{BA}_j)^2    
\end{align*}
\noindent \HighestPointX\ targets the mode, so its MSE relative to the median includes the error caused by using the mode to approximate the median: 
\begin{align*}
    \text{MSE}(\HighestPointX) = (\theta_\texttt{mode} - \theta_\texttt{median} + \beta^\texttt{HP}_j)^2 + (\sigma^\texttt{HP}_j)^2 
\end{align*}
\noindent Note that $\beta^\texttt{HP}_j$ and $\sigma^\texttt{HP}_j$ are not re-estimated, as they are already learned from~\Cref{sec:task2}. The last step is to combine these two estimates using inverse-MSE weighting to define $w$:
\begin{align*}
    \eqnmarkbox[deepgreen]{node1}{w} = \frac{\text{MSE}(\BisectArea)^{-1}}{\text{MSE}(\BisectArea)^{-1} + \text{MSE}(\HighestPointX)^{-1}}
\end{align*}
Modeling \HighestPointX\ as drawn from a Normal distribution allows for a closed-form likelihood function within the sensor fusion framework, simplifying modeling. Importantly, posterior predictive checks for a Gaussian fit to the \HighestPointX\ data (\Cref{sec:task2-gaussian}) still indicate a reasonable model fit, suggesting this simplification retains adequate descriptive power for its role in the combined model.
\begin{figure}[H]
    \centering
    \includegraphics[width=0.9\linewidth]{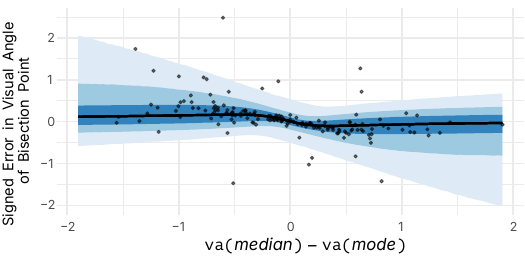}
    \caption{Posterior density, 50\%, 85\% and 95\% quantile interval of signed error of \textbf{\texttt{BAHP}} overlaid on observed signed error from participants.}
    \label{fig:area-tasks}
\end{figure}
This adaptive weighting between operators appears to better capture the observed pattern of responses compared to the simple mixture model, as confirmed by our posterior predictive checks (see \Cref{fig:task1-posterior-checks} in \Cref{sec:pp-highestpointY}). While the mixture model failed to account for the systematic change in variance of participants' judgments across different skewness conditions, the weighted average model successfully captures this change, providing a superior fit to the empirical data (see detailed posterior predictive checks in \Cref{sec:pp-bisectArea} (a) and (b)). The sensor fusion model also provides a more cognitively plausible \chreplaced{account consistent with anchoring-and-adjustment processes}{explanation for participants' behavior}---when the median and mode are close, participants tend to rely more heavily on their estimate of the mode (\HighestPointX\ operator), which is more precise and has smaller variance (\Cref{fig:area-tasks}). However, as the distribution becomes more skewed and the distance between median and mode increases, participants can no longer effectively use the mode as a heuristic. In these cases, they increasingly incorporate the \BisectArea\ operator, which requires area judgment and is thus \textit{less precise}. 

\begin{figure}[!htbp]
    \centering
    \includegraphics[width=.9\linewidth]{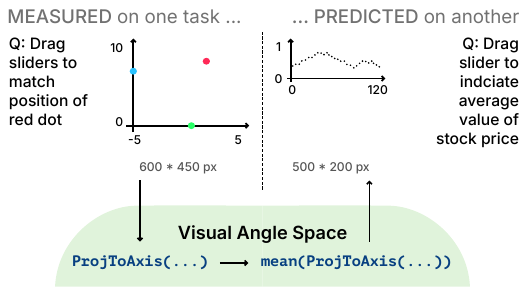}
    \caption{
    Overview of the cross-task generalization approach. Left: the \ProjToAxisY\ operator is measured by asking participants to project a dot to the y-axis on a $600\times450$ px chart. Right: the same operator, without refitting, is composed under candidate averaging strategies to predict mean-estimation responses on Moritz et al.'s~\cite{moritz2023average} $500\times200$ px scatterplot task. All composition and prediction occurs in the \textit{perceptually-uniform} visual angle space (center).}
    \label{fig:model-diagram-sec5}
\end{figure}

\section{Applying Learned Operators}
\label{sec:apply-operator}

To demonstrate the capabilities of our framework, we re-implemented experiment 2 from Moritz et al.~\cite{moritz2023average}, which asked participants to estimate the mean $y$-value of a scatterplot containing 60 data points under varying display conditions. This task requires participants to compose multiple perceptual operations---projecting individual points then aggregating across them---making it a natural test case for our operator framework. As an overview, we first re-administer \ProjToAxisY\ task from our first study to a new set of participants, deriving decoding operators for each. We then use these per-participant operators to generate predictions under three pre-registered composition strategies for a mean estimation task, and evaluate which strategy best accounts for empirical performance. Our analysis plan is pre-registered at \href{https://osf.io/wkvbe/overview}{\texttt{https://osf.io/wkvbe/}}. 

\subsection{Moritz et al. Experiment 2 ~\cite{moritz2023average}}

Moritz et al.~\cite{moritz2023average} investigated what they term \textit{variability overweighting}: the hypothesis that when estimating the average of a time series, viewers' judgments are systematically pulled toward regions of higher variability. In their second experiment ($N=420$), they manipulate the mark type between subjects, comparing line graphs to two point-based encodings: points equally spaced along the $x$-axis (\texttt{point}) and points equally spaced along the arc of the line (\texttt{pointArc}). 

The experiment used a mixed design. Between subjects, participants were randomly assigned to one of three graph types. Within-subjects factors were 2 variability levels (\chreplaced{low vs high}{0 vs 0.4}) $\times$ 2 variability positions (upper vs lower half of the chart) $\times$ $12$ random seeds (i.e.\chadded{,} stimuli datasets) $= 48$ trials per participant. Stimuli were synthetic time series of 60 data points generated from geometric Brownian motion, with noise linearly interpolated as a function of the base value. Each participant estimated the average of 48 graphs by adjusting a draggable horizontal line.

\subsection{Our Experiment Design}

\subsubsection{Tasks, Instructions, \& Training}

We first follow \Cref{sec:projection-task} to learn \ProjToAxis, and then follow the mean estimation setup in Moritz et al.~\cite{moritz2023average}. We added two training tasks for the Moritz experiment to familiarize participants with the different task interface and instructions. 

\subsubsection{Stimuli \& Apparatus}

We reused the experiment interface for \ProjToAxis\ from the first experiment, and now use uniform random sampling from $[0, 200] \times [0, 500]$ to place the dot to be projected to the $x$ and $y$-axis. For the second part of the experiment, we followed Moritz et al.'s experiment 2 stimuli design and utilized the exact same stimuli obtained from their \href{https://osf.io/aupbk/}{\faExternalLinkSquare~code repository}. We include only the \texttt{point} and \texttt{pointArc} conditions from Moritz et al., omitting the \texttt{line} condition (\href{http://mika-long.github.io/vis-decode/vis-decode-retrieve-value}{\faExternalLinkSquare~example interface}). Our current operator vocabulary includes projection operators for dot-to-axis mappings but not for line-to-value extraction, which would require an additional decomposition step (e.g., line to point) that is outside the scope of this work.

\subsubsection{Experiment Procedure \& Participants}

Similar to the first experiment (\Cref{sec:exp1-procedure}), participants first consent, then perform calibration tasks. The first part of the experiment is a short video clip explaining the \ProjToAxis\ task, two training trials with fixed stimuli across participants, then ten testing trials with randomized stimuli. Then the participant is randomly assigned to one of two conditions, \texttt{point} or \texttt{pointArc}, and each condition has two fixed training trials and 48 testing trials, presented to them in random order. In total, each participant performed $10 + 48 = 58$ trials. Finally, participants provided demographic information and optional text feedback on their strategies and the clarity of instructions.

Similar to the first experiment, we recruited 20 participants in an IRB-approved study \chadded{(\texttt{STU00223203})}\chdeleted{[omitted for anonymization]} on \texttt{Prolific.co} \chadded{using the same eligibility criteria, }\chdeleted{who are between the ages of 18 and 65, have normal or corrected-to-normal vision, speak fluent English, and reside in the United States. The study was }distributed to a gender-balanced sample\chdeleted{ of participants}. \chadded{No participant met the exclusion criteria, and participants ranged in age from 20 to 62 (M = 38.2, SD = 14.5).} We paid each participant \$3.10 for completing the experiment, with a potential bonus of up to \$2.90 depending on the participants' precision. The experiment took on average 20.5 minutes to finish, resulting in an hourly wage of \$17.56 / hr\chdeleted{, exceeding the US federal minimum wage}.

\subsection{Analysis Approach}

Similar to \Cref{sec:exp1-analysis}, we used multi-level Bayesian regression to analyze our results. We fit Bayesian hierarchical mixed effects model on data from part 1 for \ProjToAxisY, with all parameters achieving bulk effective sample size (\texttt{Bulk\_ESS}) and tail effective sample size (\texttt{Tail\_ESS}) values ranging from $2981$ to $7981$.
We utilized the same set of \texttt{R} packages as \Cref{sec:exp1-analysis}, with the addition of \texttt{jsonlite}~\cite{jsonlite} and \texttt{purrr}~\cite{purrr}. 

\subsubsection{Hypothesized Strategies}

Each trial presents 60 points $(x_i, y_i)$, $i \in [60]$ where the participant estimates the mean $y$-value, which we denote $\hatmuY$. We posit strategies for mean estimation composed of projection operations and mental averaging.
Our strategies vary along two dimensions: \textit{projection path} (project-once vs. project-twice) and \textit{aggregation rule} (mean, median, sensor-fusion), yielding six candidate strategies.\footnote{Our pre-registered analysis specified three strategies (project-once $\times$ mean, project-once $\times$ median, project-twice $\times$ sensor-fusion). During analysis, we recognized these three span a partial crossing of two orthogonal design factors, projection path and aggregation rule, and completed the full $2 \times 3$ factorial to avoid selectively pairing projection paths with aggregation strategies. No data-dependent decisions guided which cells to add; the posterior prediction pipeline is deterministic given the learned operator and the composition rule.}
 
\noindent \textbf{\textsf{Projection paths.}} We consider two possible projection paths: 
\begin{enumerate}
    \item \textbf{Project-once.} The participant projects each point directly to the $y$-axis then takes an average:
    \[
        \hatmuY = \vaY^{-1}\Bigl[
         \Average\Bigl(
          \ProjToAxisY\bigl(
           \vaY(y_i), \vaX(x_i)
          \bigr)
         \Bigr)
        \Bigr]
    \]
    where $\Average$ represents one of the aggregation rules (below). 
    As each projection traverses a horizontal distance of $\vaX(x_i)$, points farther from the y-axis incur more noise. This may not be a plausible account, motivating our second projection path.
 
    \item \textbf{Project-twice.} Instead of projecting each point all the way to the y-axis, the participant first projects each point to an imagined vertical reference line at the midpoint (in this case, $x = 60$):
    \[\hat \theta_i = \ProjToAxisY \bigl(
      \vaY(y_i), \vaX(|x_i - 60|)
    \bigr)\]
    The participant then averages the projected points and projects that average from the midpoint to the y-axis:
    \begin{equation*}
        \hatmuY = \vaY^{-1}\Bigl[
         \ProjToAxisY\bigl(
          \Average(\hat\theta_i), \vaX(60)
         \bigr)
        \Bigr]
    \end{equation*}
    This account may be more plausible if participants visually estimate the mean around the center of the data along both axes rather than at one edge.
\end{enumerate}

\noindent \textbf{\textsf{Aggregation rules.}} Given internally-estimated visual angles $\hat\theta_i$ for each $y$-value $y_i$ (from either projection path), we consider three possible definitions of the perceptual $\Average$ operation:
 
\begin{enumerate}[nosep]
    \item \textbf{Mean:} $\Average(\hat\theta_i) = 
    \mathrm{mean}_{i \in [60]}(\hat\theta_i)$
 
    \item \textbf{Median:} $\Average(\hat\theta_i) = 
    \mathrm{median}_{i \in [60]}(\hat\theta_i)$
 
    \item \textbf{Weighted mean:} 
    $\Average(\hat\theta_i) = 
    \sum_{i \in [60]} w_i \, \hat\theta_i$, where 
    $w_i = \frac{\mathrm{MSE}(\hat\theta_i)^{-1}}
    {\sum_{k \in [60]} \mathrm{MSE}(\hat\theta_k)^{-1}}$ 
    weights each point by the inverse MSE of the \ProjToAxisY\ operation that produced it. 
\end{enumerate}

\subsection{Results}
\subsubsection{Part 1: \ProjToAxisY\ Operator Generalizes Across Participant Groups}
\label{sec:part2-part1}

\begin{wrapfigure}[21]{r}{0.45\linewidth}
    \centering
    \vspace{-1.5em}
    \includegraphics[width=\linewidth]{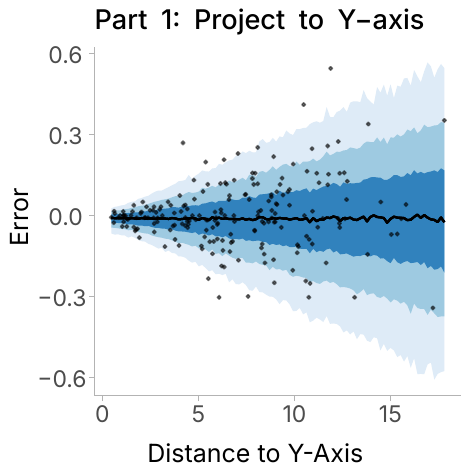}
    \caption{Error vs. distance between start and end points of projection for \ProjToAxisY, measured in radian units, where points show observed errors with blue bands representing 50\%, 80\%, and 95\% posterior predictive intervals, marginalizing over participants.}
    \label{fig:part2-task1-project-y}
\end{wrapfigure}

For the first part of the experiment, we fit the same model as in \Cref{sec:projection-task} and obtained a similar trend to \Cref{fig:projection-overall} (c): error increases with the visual angle distance between the point and the $y$-axis (\Cref{fig:part2-task1-project-y}; posterior predictive checks in \Cref{sec:pp-check-exp2}). This provides preliminary evidence that our model is capturing a trend that generalizes across participant pools. 

\subsubsection{Part 2: Comparing Composition Strategies}

\begin{figure}[H]
    \centering
    \includegraphics[width=\linewidth]{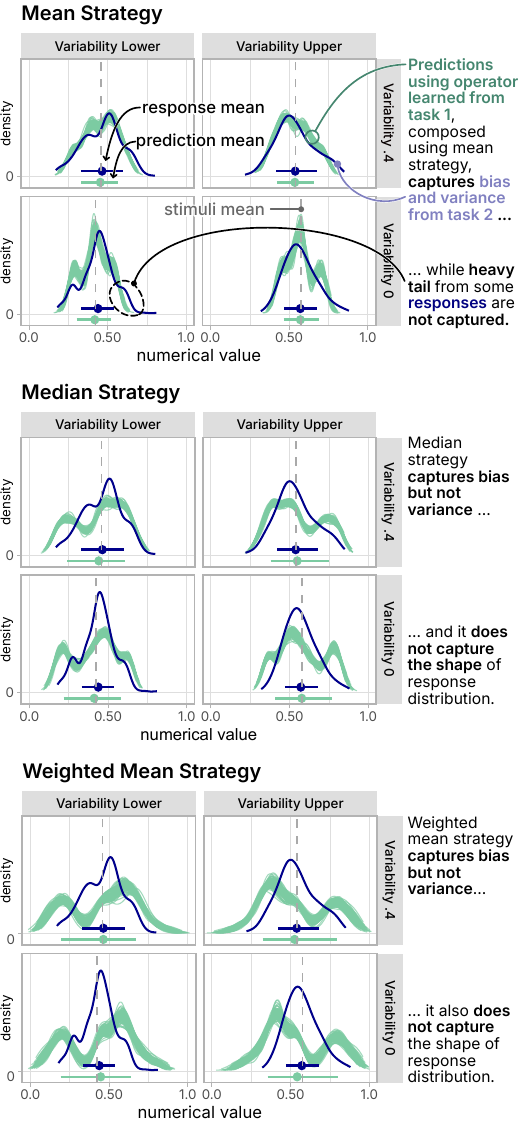}
    \caption{Cross-task posterior predictive density overlay plot, each with 100 draws, faceted by variability upper/lower vs. variability 0/0.4 experiment conditions. Point intervals indicate the mean and 68\% quantile interval (approximately $\pm$ 1 SD) for observed responses and model predictions respectively.}
    \label{fig:part2-KDE}
\end{figure}

The model fitted in \Cref{sec:part2-part1} gives us posterior distributions for the parameters of each participant's \ProjToAxisY\ operator. Using these parameters, we can generate a predictive distribution for how each participant would respond if they had used each combination of \textit{projection path} $\times$ \textit{aggregation rule} on each mean-estimation trial in part 2 (Moritz et al.~\cite{moritz2023average} experiment 2). We emphasize that these posterior predictions are generated entirely from operators learned on the \ProjToAxisY\ task on charts with a different size and aspect ratio, \textit{then} re-composed under each prospective aggregation strategy to create predictions. \textit{No parameters are fit to the mean-estimation data} (\Cref{fig:part2-KDE}). 

\chadded{Operator-level uncertainty is propagated through composition by Monte Carlo (MC) on the joint posterior of \ProjToAxisY's parameters: for each posterior draw, we sample one projected value per stimulus point, apply the candidate aggregation rule (and, for \textbf{project-twice}, a second sampled projection), and accumulate the resulting predictive distribution. We do not claim this composition reflects what viewers internally compute. For the 60-point stimuli used here, this procedure is exact up to MC error, and for substantially larger object counts, analytical approximations to the aggregate variance would be a sensible efficiency improvement.}

Figure~\ref{fig:part2-KDE} shows posterior predictive density overlays for all three strategies for \textbf{project-twice}, faceted by variability conditions. The project-twice mean strategy (\Cref{fig:part2-KDE}, top) captures both the central tendency and spread of observed responses across all four conditions. The median strategy (\Cref{fig:part2-KDE}, middle) matches observed bias but produces prediction intervals that are too wide. The weighted-mean strategy (\Cref{fig:part2-KDE}, bottom) shows a similar pattern. We defer posterior predictive density overlays for the \textbf{project-once} strategies to~\Cref{sec:posterior_1proj}, as these produced substantially narrower prediction intervals that failed to match observed responses. 

\subsubsection{Implications}

That the project-twice mean strategy can capture bias and variance on a structurally different task, where five alternatives fail, has two implications. First, it provides evidence that individually-measured operators can generalize through composition to predict performance on unseen charts with a different size and aspect ratio, \textit{without refitting}. Together with the operator-level results in \Cref{sec:operator_results}, this constitutes an existence proof that visual decoding operators can be measured in isolation, composed under candidate strategies, and evaluated against held-out task performance. The strategy comparison also allows us to rule out some prospective perceptual aggregation functions: for example, it seems unlikely people are mentally calculating medians here.

\section{Discussion}

Together, \Cref{sec:build-operator} and \Cref{sec:apply-operator} establish that operators can be isolated, parameterized, and composed to predict task performance on unseen display conditions. Below, we consider \chadded{what it takes to build an operator library and what such a library would enable}\chdeleted{how operators provide a compositional interface for formalizing graphical perception knowledge and enabling a new class of design tools} (\Cref{sec:discussion-dsl}) \chadded{and} what kind of theoretical contribution \chadded{our approach}\chdeleted{this} represents within existing frameworks for model evaluation (\Cref{sec:discussion-process})\chdeleted{, and what empirical work is needed to build out an operator library (\mbox{\Cref{sec:futurework})}}.

\subsection{\chadded{Toward an Operator Library for Graphical Perception}\chdeleted{A Compositional Interface to Formalize Graphical Perception Knowledge}}
\label{sec:discussion-dsl}

\chadded{\textbf{What makes an operator valid?} A visual decoding operator is minimally valid when it satisfies four necessary conditions: 1) it can be isolated by a task instruction, 2) it can be measured in a perceptually uniform space, 3) its bias and variance parameters can be estimated from that isolated task, and 4) the same operator appears recognizably in decompositions of multiple chart~$\times$~task combinations. These conditions guarantee that an operator is measurable and reusable, but not that it is useful. A new operator is warranted when the decomposition of a task requires a perceptual act that cannot be expressed as a composition of existing valid operators. A perfectly valid operator could still be perceptually unmotivated or empirically weak. Settling on a shared set of desirable properties, such as predictive transfer to untested chart~$\times$~task pairs, grounding in known capabilities of the visual system, or parsimony, is a question we think the community is better positioned to answer collectively than any single paper. We see the above four necessary conditions as the floor, not the ceiling.}

\chadded{\textbf{The work ahead.} The five operators we characterized are not intended to be exhaustive; they are an existence and feasibility proof scoped to some tasks on Cartesian charts utilizing position encodings. Operators for other common perceptual acts, such as reading off a non-Cartesian coordinate, tracing a path across a visual field, or extracting and interpolating values from colors, remain uncharacterized. Such operators would need to be isolated and measured to extend this framework to the full range of tasks visualization researchers study. Further, tasks that require particular visualization literacy~\cite{lily2025autoethnography, nobre2024reading} or decision making~\cite{sarma2025more,kaleVisualReasoningStrategies2021,wu2023rational} skills in addition to perceptual capabilities are not modeled nor captured by the current framework. Visualizations that use position non-metrically (e.g., in network diagrams) also fall outside the current scope, as the projection operators characterized here do not capture adjacency or grouping. All of these areas represent possible directions for building out the library of operators. For example, a natural next step might be to build an operator for axis reading, which would compose directly with our projection operators: projection outputs an angular position, and an axis reading operator could be fitted to map that position to a numerical value on a scale given the angular positions of reference marks. Existing works, such as the cyclical power model~\cite{hollands2000bias} and slider decoration effects~\cite{matejka2016effect}, offer starting points. Reformulating either model as an operator would extend predictions to charts where axis interpolation, or the elicitation interface itself, carries non-trivial error.}

\chadded{\textbf{What a fuller library would enable.} } Operators \chadded{could} provide \chreplaced{an interface that ranking-based frameworks cannot}{a different kind of interface}. Because operators are \chadded{designed to be }parameterized by \textit{perceptual units}\footnote{here visual angle, but we hope by other perceptual units in the future, such as values in perceptually-uniform color spaces.} rather than screen pixels, they generalize across display context. And because operators are \chadded{designed to be} composable, complex tasks can be expressed as sequences of operations, each carrying a quantified cost. This composability opens several practical applications in visualization design. For instance, operators can serve as cost functions in a \textit{query planner}: given a dataset and an analytic goal, the planner enumerates candidate visualization designs, decomposes each into its operator sequence, and predicts a user's likely decoding path by balancing the cognitive cost of each operation against its expected informational gain. This differs from current ranking-based recommendation systems, in that the predicted error distribution tells the designer not just which option is better but \textit{by how much} and \textit{with what uncertainty}. \chadded{Further, once operator-level error is established as a perceptual baseline, remaining error can be more readily attributed to non-perceptual sources such as task mistranslation, training effects, or elicitation artifacts~\mbox{\cite{sarma2024tasks}}.} The error measurements in this paper provide initial cost estimates \chadded{of five operators}; a fuller library \chdeleted{(\mbox{\Cref{sec:futurework}})} would expand the space of tasks and chart types over which such a system can reason. \chadded{The experiments required to populate such a library constitute a research program comparable in scope to the channel-ranking studies that followed Cleveland and McGill~\cite{cleveland1984graphical}, reframed around reusable primitives whose compositions generate predictions beyond the conditions they were measured in.}

\subsection{Toward Algorithmic-Level Theories of Visualization Interpretation}
\label{sec:discussion-process}

Visual decoding operators specify \textit{how} a person extracts a piece of information from a chart: which visual features serve as input, what transformation is applied, and what error distribution results. This is what distinguishes the framework from a behavioral summary. A behavioral summary records that viewers underestimate means in scatterplots with high variability; an operator \chdeleted{account }specifies the projection and averaging operations that produce the underestimate, predicts its magnitude, and identifies which component to revise when the prediction fails.

This distinction maps onto a well-known framework for evaluating models of information-processing systems. Marr~\cite{marr2010vision} proposed three levels of analysis: the \textbf{computational level} (\textit{what problem is the system solving?}), the \textbf{algorithmic level} (\textit{what procedures does it use?}), and the \textbf{implementational level} (\textit{how is it physically realized?}). Most modeling work in visualization operates at the computational level --- we have extensive catalogs of tasks and methods for measuring their aggregate error~\cite{elliottDesignSpaceVision2021}, and recent models simulate observable behavior such as eye scan paths~\cite{shi2025chartist} or leverage LLMs to predict task outcomes~\cite{10681139}. These approaches describe what the system does, or predict what it will do, but not \textit{how}. Visual decoding operators engage at the \textit{algorithmic level}: they specify the sequence of transformations a viewer executes and the error each transformation introduces. 

The value of this level of description extends beyond taxonomy. Wichmann and Geirhos~\cite{wichmann2023deep} argued that model evaluation is multidimensional: a good model should both predict and explain. Our cross-task results in \Cref{sec:apply-operator} demonstrate prediction, in that operators learned on one task generate calibrated posterior predictive distributions for a structurally different task. The sensor fusion finding in \Cref{sec:Task1} \chreplaced{illustrates the framework's potential on the explanation side:}{demonstrates explanation, in that} the weighted combination of \HighestPointX\ and \BisectArea\ not only fits the data better than a mixture model but \chreplaced{is compatible with an anchoring-and-adjustment account}{provides a mechanistic account} of why variance changes with skewness, an account that a purely descriptive model would not generate.

Van Rooij et al.~\cite{van2024makes} have argued that psychology and its adjacent fields remain dominated by what they term \textit{naive empiricism}: accumulating experimental results without theoretical infrastructure to organize them into explanatory models. The visualization community faces a version of the same challenge. Verbal theories in graphical perception often provide post-hoc accounts of empirical patterns that appear explanatory but lack the computational structure to generate quantitative predictions for new conditions. The operator framework is one response to this challenge: it provides units whose predictions are falsifiable at the component level, so that when a composed prediction fails, the failure points to which operator needs revision rather than leaving an undifferentiated residual.

\section{Conclusion}
This work introduced visual decoding operators as a reusable unit of analysis for modeling how people interpret visualizations. We showed that common tasks decompose into constituent perceptual operations whose precision and variability can be characterized empirically through hierarchical modeling. Composing these operators under candidate strategies yields generative models that produce testable predictions of task performance. In a pre-registered cross-task evaluation, operators derived from a projection task generated posterior predictive distributions that captured participants' mean estimation behavior under unseen display conditions, providing initial evidence that individual operators can generalize beyond the tasks used to measure them. \chreplaced{Building out a full library of visual decoding operators is a substantial empirical undertaking, one that is beyond the scope of a single paper. Nevertheless, this operator-based approach demonstrates the viability of a new way of doing empirical visualization research.}{Because each operator carries a quantified error profile, the framework provides the primitives for design tools that can predict the perceptual cost of a decoding path and compare candidate visualizations on that basis.}

\section*{Supplemental Materials}
\label{sec:supplemental_materials}
All supplemental materials are available on OSF at \href{https://osf.io/prtfq/overview}{\texttt{https://osf.io/prtfq}}, released under a \href{https://creativecommons.org/licenses/by/4.0/}{\ccby{} CC BY 4.0 license}.
In particular, they include (1) data collected from the first and second experiment, as well as stimuli data from experiment of Moritz et al.~\cite{moritz2023average} obtained via \url{ https://osf.io/aupbk/}, (2) experiment analysis scripts in \texttt{R}, \texttt{stan}, and quarto, (3) the corresponding generated html files for the analysis scripts, (4) figures used in this paper, and (5) a full version of this paper with all appendices. 

The first experiment interface can be found at \url{https://mika-long.github.io/vis-decode/vis-decode-slider} and the second experiment interface can be found at \url{https://mika-long.github.io/vis-decode/vis-decode-retrieve-value}. 

\section*{Figure Credits and Copyrights}
\label{sec:figure_credits}

All figures in this paper is created by the authors and thus remain under the authors' own copyright, with the permission to be used here. We also share them under a \href{https://creativecommons.org/licenses/by/4.0/}{\ccby{} CC BY 4.0 license} at \href{https://osf.io/prtfq/overview}{\texttt{https://osf.io/prtfq}}.

\acknowledgments{%
This research was supported in part by NSF \#2211939, \#2106197, \#2103794, \#2312991, as well as DAPLab corporate support in the form of funding and/or compute from Amazon, IntellectAI, Infosys, Tidalwave, Veris, Shopify, Microsoft, Thinking Machines, Dandy, Perplexity, and Daytona. The views and conclusions presented here are those of the authors and should not be interpreted as representing the official positions of the funding organizations. In addition, the authors wish to thank members of the MU Collective Lab, Lily Ge, Mandy Cai, and Charlotte Li at Northwestern University and members of the ReVISit team, Alex Lex, Lane Harrison, Andrew McNutt, Zach Cutler, Yiren Ding, Jay Kim, and Jack Wilburn for providing technical support and feedback.}

\bibliographystyle{abbrv-doi-hyperref-narrow}

\bibliography{00-references}

\appendix 
\crefalias{section}{appendix} 
\newpage


\section{Slider Interface}
\label{sec:slider-interface}

\begin{wrapfigure}[23]{r}{0.45 \linewidth}
    \vspace{-1em}
    \centering
    \includegraphics[width=\linewidth]{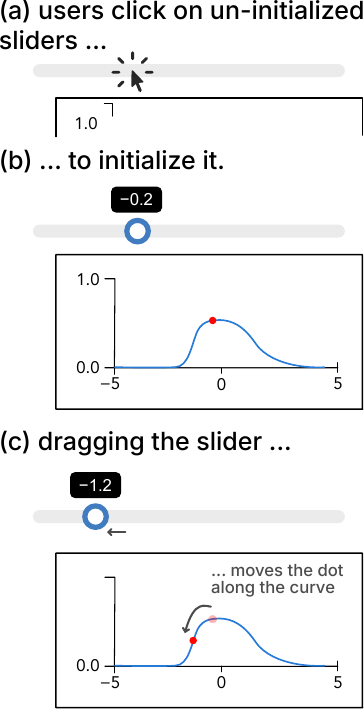}
    \label{fig:slider_demo}
    \vspace{-1em}
\end{wrapfigure}

\chadded{To minimize potential confounds, we implemented a custom elicitation interface --- a slider with no default starting position to prevent anchoring effects. The slider was positioned away from the chart axes to prevent participants from directly aligning the slider position with axis values, encouraging them to focus on the red dot annotation that appears on the curve itself. Participants first clicked anywhere on the slider to initialize it (\textsf{a}), which displayed a red dot annotation on the curve at the corresponding position (\textsf{b}). They could then control the red dot's position by dragging the slider or using left and right arrow keys (\textsf{c}). We chose this slider-based approach over alternatives, such as direct clicking on the curve, because it eliminated ambiguity about the source of error while constraining participants to perform the specific visual operators under investigation.} 

\section{Full Hierarchical Model Specifications}
\label{sec:full-spec}

We follow the hierarchical notation system of Gelman and Hill~\cite{gelman2007data} and denote participants' selection on the X-Axis and Y-Axis with $\hat{x}, \hat{y}$ respectively and the true answer with $x, y$. Unless otherwise specified, standard deviation parameters ($\sigma$) follow $\mathcal{N}^+(0, 1)$ and other parameters follow $\mathcal{N}(0, 1)$ as (hyper-)prior distributions. The full model specifications are in the quarto and html files in supplemental materials hosted at \href{https://osf.io/prtfq/overview?view_only=1c7104a9488940a6aa5a042d41bb1232}{\texttt{https://osf.io/prtfq}}. 

\section{Screenshot of Experiment Interface}
\label{sec:experiment-interface}

\begin{figure}[!hbtp]
  \centering
    \includegraphics[width=\linewidth]{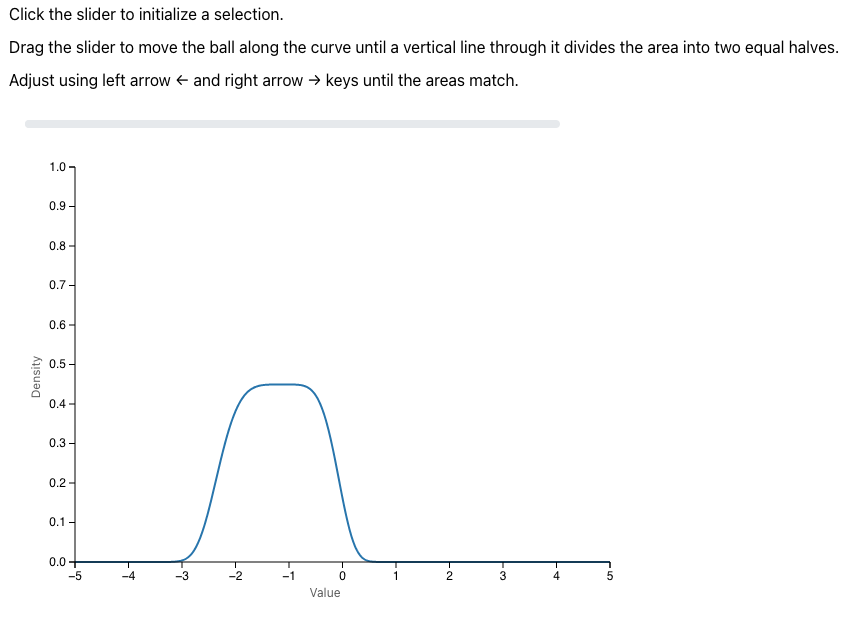}
    \caption{An example of the experiment interface. The generalized skewed-$t$ distribution is parameterized with $\mu = -1.1, \sigma = 0.7, \lambda = -0.1, p = 4.6, q = 31.7$.}
    \label{fig:example-dist} 
\end{figure}

\section{Skewed Generalized T Distribution}
\label{sec:skew-t-pdf}

The skewed generalized $t$ distribution (SGT), introduced by Theodossiou~\cite{theodossiouFinancialDataSkewed1998}, is a univariate, unimodal distribution. We chose it as our data-generating distribution for its flexibility, minimal distributional parameters, and capacity to model a wider range of data characteristics than Gaussian, Laplace, and many other common distributions. Specifically, the SGT allows independent control over skewness (asymmetry), kurtosis (tail heaviness), and scale, enabling us to generate distributions that range from highly symmetric to strongly skewed, and from light-tailed (platykurtic) to heavy-tailed (leptokurtic). This flexibility is crucial for testing human perception across diverse distributional shapes that participants might encounter in real-world data visualization contexts.

The SGT probability density function is defined as

\begin{align} \label{eq:1}
f_{\text{SGT}}(x; \mu, \sigma, \lambda, p, q) = \\ 
\frac{p}{2\,v\sigma\, q^{1/p} B(1/p, q) \left[ 1 + \frac{|x - \mu|^p}{q (v\sigma)^p(1 + \lambda \cdot \text{sign}(x-\mu))^p}  \right]^{q + 1/p}}
\end{align} 

where $\mu$ is \textit{mode}, $\sigma$ is \textit{scale}, $\lambda$ is \textit{skewness}, and $p$ and $q$ are \textit{kurtosis} parameters. $B(\cdot, \cdot)$ is the beta function and $v$ is a variance adjustment parameter, defined as \[v=\left(q^{-\frac{1}{p}} \left[(3 \lambda^2 + 1) \frac{B(\frac{3}{p}, q-\frac{2}{p})}{B(\frac{1}{p}, q)} - 4 \lambda^2 \left(\frac{B(\frac{2}{p}, q-\frac{1}{p})}{B(\frac{1}{p}, q)}\right)^2\right] 
 \right)^{-\frac{1}{2}}\]

 To ensure sufficient variation in the visualized data range of $x \in [-5, 5], y \in [0, 1]$, we sampled the values for the parameters as $\mu \sim \text{Uniform}[-2, 2]$, $\sigma \sim \text{Uniform}[0.5, 2.5]$, $\lambda \sim \mathcal{N}(0, 0.33)$, $p \sim \text{Uniform}[2, 4]$, and $q \sim \text{Uniform}[1, 50]$. 



\section{Additional posterior check related plots}
\label{sec:additional-posterior-checks}

\subsection{Checks for Experiment 1}

Following best practices, we plot and examine the posterior predictive check plots and the Probability Integral Transform-Empirical Cumulative Distribution Function (PIT-ECDF) calibration plots. Throughout our results, posterior predictive draws recover the shape of the participants' responses. For details on how to interpret posterior predictive plots, refer to Gabry et al~\cite{gabry2019visualization}.

\subsubsection{Remaining Two Projection Operators}

\begin{figure}[H]
    \centering
    \hfill
    \begin{subfigure}[b]{0.6\linewidth}
        \centering
        \includegraphics[width = \linewidth]{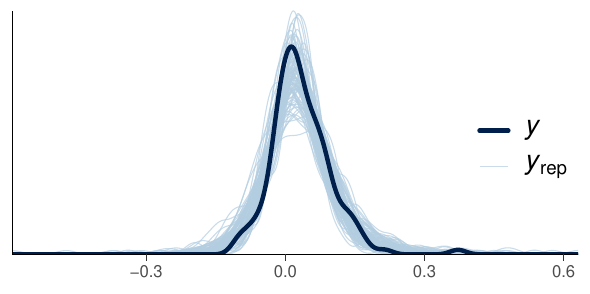}
        \caption{}
        \label{fig:task5-x-pp-check}
    \end{subfigure}
    \hfill
    \begin{subfigure}[b]{0.35\linewidth}
        \centering
        \includegraphics[width = \linewidth]{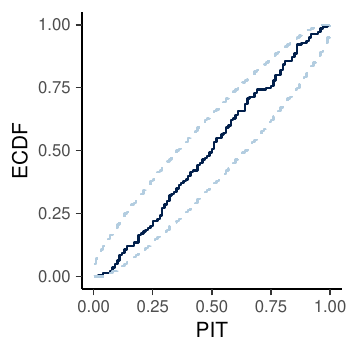}
        \caption{}
        \label{fig:task5-x-pit-ecdf}
    \end{subfigure}
    \begin{subfigure}[b]{0.6\linewidth}
        \centering
        \includegraphics[width=\textwidth]{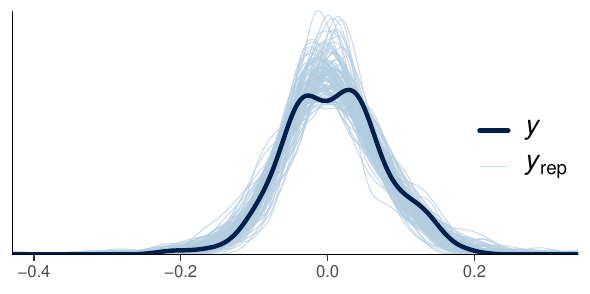}
        \caption{}
        \label{fig:task5-y-pp-check}
    \end{subfigure}
    \begin{subfigure}[b]{0.35\linewidth}
         \centering
        \includegraphics[width=\textwidth]{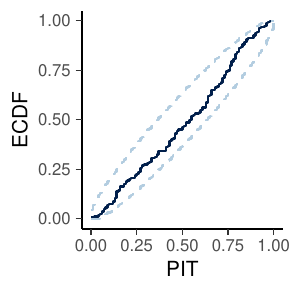}
        \caption{}
        \label{fig:task5-y-pit-ecdf}
    \end{subfigure}
    \caption{Plots for \ProjToAxisX\ and \ProjToAxisY: (a), (c) Posterior predictive checks with 100 draws for $x$-axis and $y$-Axis respectively. (b), (d) PIT-ECDF calibration plot for $x$-axis and $y$-axis respectively.}
    \label{fig:placeholder}
\end{figure}

\subsubsection{\HighestPointY\ Operator}
\label{sec:pp-highestpointY}

\begin{figure}[!htbp]
    \centering
    \begin{subfigure}[b]{0.6\linewidth}
        \centering
        \includegraphics[width = \textwidth]{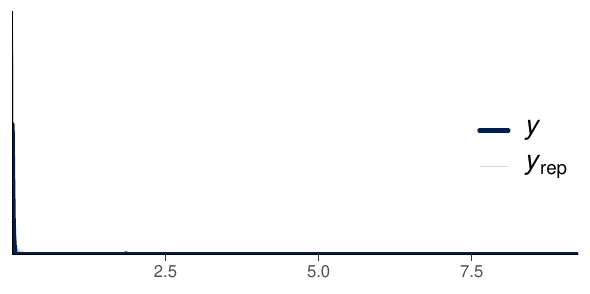}
        \caption{}
        \label{fig:task2-y-ppcheck}
    \end{subfigure}
    \hfill
    \begin{subfigure}[b]{0.35\linewidth}
            \centering
    \includegraphics[width = \textwidth]{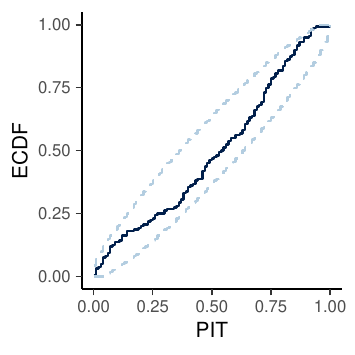}
    \caption{}
    \label{fig:task2-y-pit-ecdf}
    \end{subfigure}
    \caption{Plots for \HighestPointY: (a) Posterior predictive check with 100 draws, (b) PIT-ECDF calibration.}
    \label{fig:task2-y-posterior-checks}
\end{figure}

\subsubsection{Gaussian model for \HighestPointX}
\label{sec:task2-gaussian}

The mathematical model for modeling the signed error for the highest point task (\Cref{sec:task2}) on the $x$-axis is as follows: 
\begin{align*}
    \text{error}_i &\sim \mathcal{N}(\mu_{\text{PID}[i]}, \sigma_{\text{PID}[i]}^2) \\ 
    \mu_j &\sim \mathcal{N}(\bar{\mu}, \bar{\sigma}^2), j \in [16] \\ 
    \sigma_j &\sim \mathcal{N}(\alpha, \beta^2), j \in [16] \\ 
    \bar{\mu}, \alpha &\sim \mathcal{N}(0, 1) \\ 
    \bar{\sigma}, \beta &\sim \mathcal{N}^+(0, 1) 
\end{align*}

\begin{figure}[!htbp]
    \centering
    \begin{subfigure}[b]{0.6\linewidth}
        \centering
    \includegraphics[width=\textwidth]{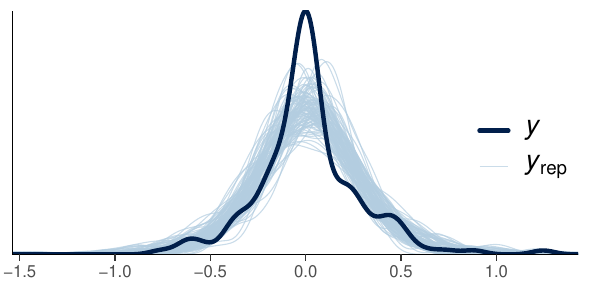}
    \caption{}
    \label{fig:task2-gaussian-pp-check}
    \end{subfigure}
    \hfill
    \begin{subfigure}[b]{0.35\linewidth}
            \centering
    \includegraphics[width=\textwidth]{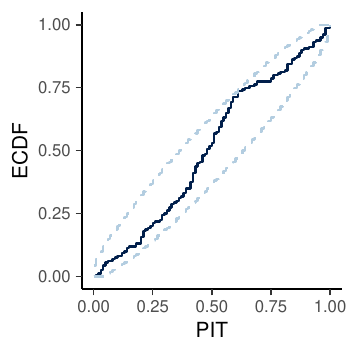}
    \caption{}
    \label{fig:task2-gaussian-pit-ecdf}
    \end{subfigure}
    \caption{Plots for \textbf{\texttt{HighestPoint}}\textsubscript{X} task for \textbf{error on the X-Axis}\chdeleted{, except here we are using a Normal distribution instead of a Laplace distribution to model}: (a) Posterior predictive check with 100 draws, (b) PIT-ECDF calibration chart. }
    \label{fig:task2-x-gaussian}
\end{figure}

Note that the Gaussian model does not quite capture the response distribution, but the good PIT-ECDF performance tells us that it's still a relatively good fit, and thus it works for the \BisectArea\ operator in \Cref{sec:Task1}. 


\subsubsection{Max Slope Operator}

\begin{figure}[H]
    \centering
    \begin{subfigure}[b]{0.6\linewidth}
        \centering
    \includegraphics[width=\textwidth]{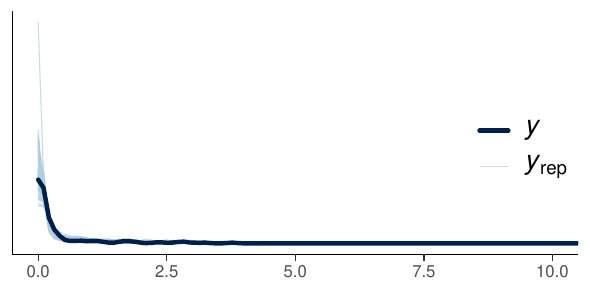}
    \caption{}
    \label{fig:task4-pp-check}
    \end{subfigure}
    \hfill
    \begin{subfigure}[b]{0.32\linewidth}
            \centering
    \includegraphics[width=\textwidth]{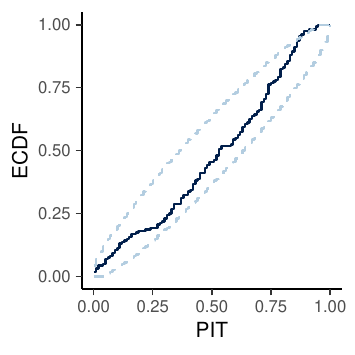}
    \caption{}
    \label{fig:task4-pit-ecdf}
    \end{subfigure}
    \caption{Plots for \MaxSlope: (a) Posterior predictive check with 100 draws, (b) PIT-ECDF calibration plot.}
    \label{fig:task4-posterior-checks}
\end{figure}

\subsubsection{Bisect Area Operator}
\label{sec:pp-bisectArea}

\begin{figure}[H]
    \centering
    \begin{subfigure}[b]{0.85\linewidth}
        \centering
    \includegraphics[width=\textwidth]{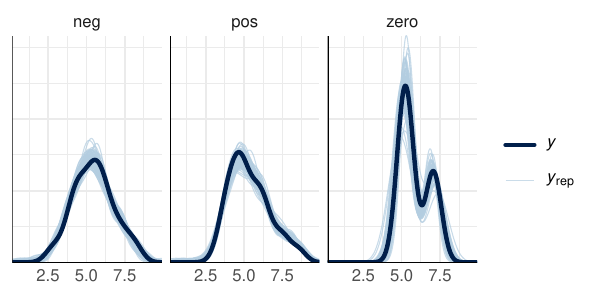}
    \caption{}
    \label{fig:task1-ppcheck}
    \end{subfigure}
    \hfill
    \begin{subfigure}[b]{0.95\linewidth}
            \centering
    \includegraphics[width=\textwidth]{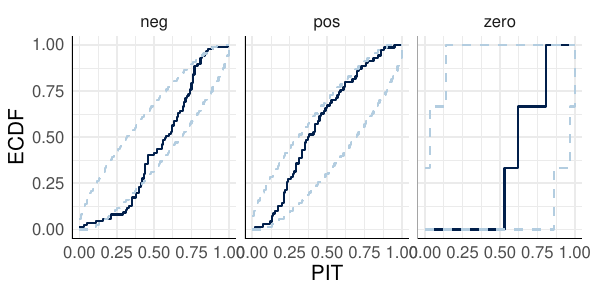}
    \caption{}
    \label{fig:Task1-pit-ecdf}
    \end{subfigure}
    \caption{Plots for \BisectArea. (a) Posterior predictive checks with 100 draws, where the plot is faceted to have three parts, where the label indicates whether the sign of $\lambda$, the skewness parameter, is negative, positive, or zero, and (b) PIT-ECDF calibration chart, where the plot is faceted to have three parts, where the label indicates whether the sign of $\lambda$, the skewness parameter, is negative, positive, or zero. (a) and (b) demonstrate good fits.}
    \label{fig:task1-posterior-checks}
\end{figure}

\subsection{Checks for Experiment 2}

\subsubsection{Part 1 -- \ProjToAxisY}
\label{sec:pp-check-exp2}

\begin{figure}[H]
    \centering
    \begin{subfigure}[b]{0.6\linewidth}
        \centering
    \includegraphics[width=\textwidth]{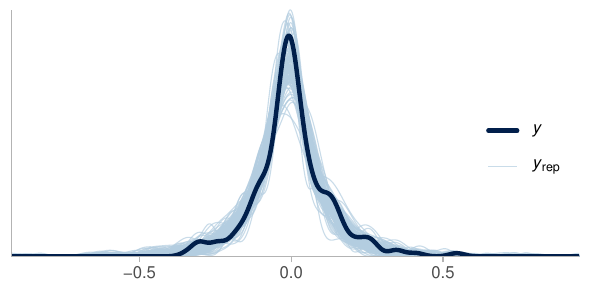}
    \caption{}
    \label{fig:part2-task1-pp-check}
    \end{subfigure}
    \hfill
    \begin{subfigure}[b]{0.35\linewidth}
            \centering
    \includegraphics[width=\textwidth]{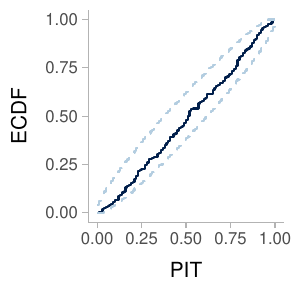}
    \caption{}
    \label{fig:part2-task1-pit-ecdf}
    \end{subfigure}
    \caption{Plots for \ProjToAxisY\ for \textbf{Experiment 2}. (a) Posterior predictive checks with 100 draws for the Y-Axis. (b) PIT-ECDF calibration plot for Y-Axis.}
    \label{fig:part2-task1}
\end{figure}

\subsubsection{Part 2 -- Project-Once}
\label{sec:posterior_1proj}

The following are all plots generated using \texttt{bayesplot}. 

\begin{figure}[H]
    \begin{subfigure}[b]{0.35\textwidth}
        \centering
        \includegraphics[width=\textwidth]{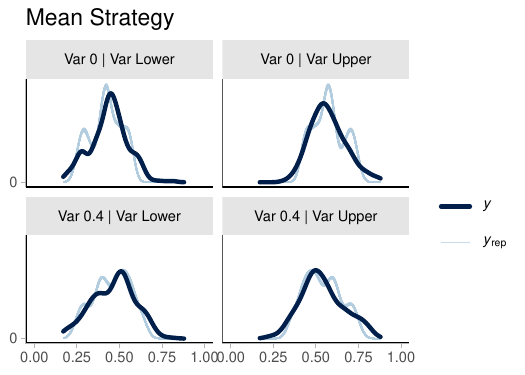}
        \caption{}
        \label{fig:1proj_mean}
    \end{subfigure}
    \hfill
    \begin{subfigure}[b]{0.35\textwidth}
        \centering
        \includegraphics[width=\textwidth]{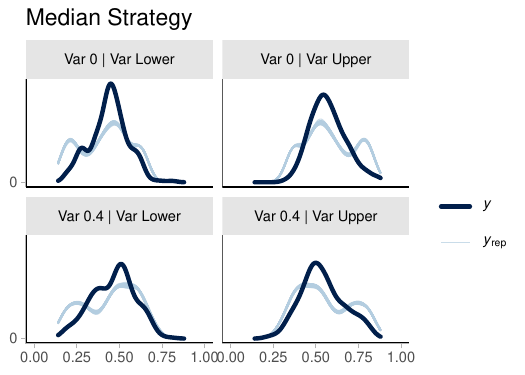}
        \caption{}
        \label{fig:1proj_median}
    \end{subfigure}
    \hfill
    \begin{subfigure}[b]{0.35\textwidth}
        \centering
        \includegraphics[width=\textwidth]{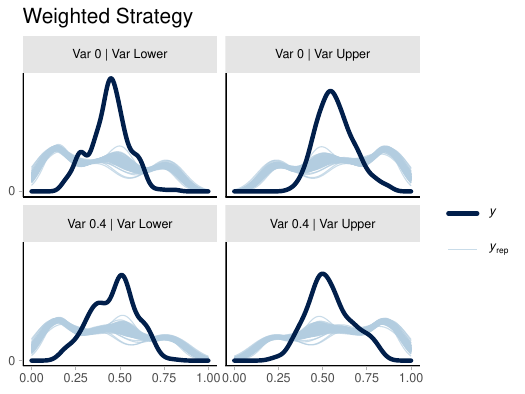}
        \caption{}
        \label{fig:1proj_weighted}
    \end{subfigure}
    \caption{\small Predictions composed under \textbf{project-once} strategy for (a) mean, (b) median, and (c) weighted.}
    \label{fig:1proj}
\end{figure}

\subsubsection{Part 2 -- Project-Twice}

The following are all plots generated using \texttt{bayesplot}. 

\begin{figure}[H]
    \begin{subfigure}[b]{0.35\textwidth}
        \centering
        \includegraphics[width=\textwidth]{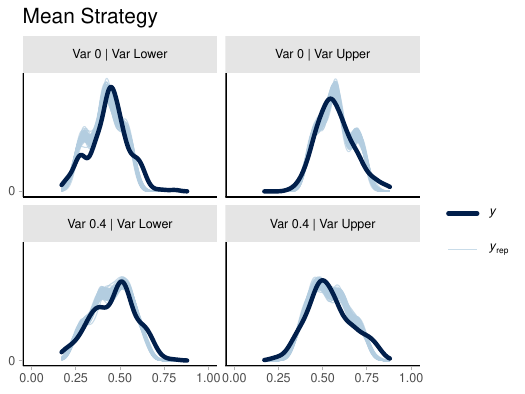}
        \caption{}
        \label{fig:2proj_mean}
    \end{subfigure}
    \hfill
    \begin{subfigure}[b]{0.35\textwidth}
        \centering
        \includegraphics[width=\textwidth]{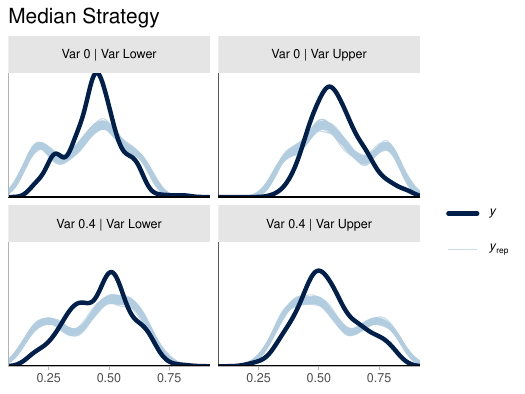}
        \caption{}
        \label{fig:2proj_median}
    \end{subfigure}
    \hfill
    \begin{subfigure}[b]{0.35\textwidth}
        \centering
        \includegraphics[width=\textwidth]{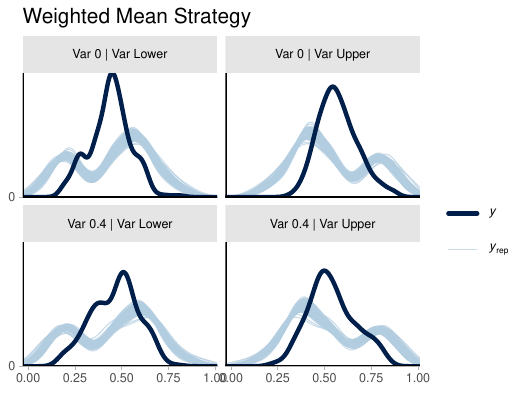}
        \caption{}
        \label{fig:2proj_weighted}
    \end{subfigure}
    \caption{\small Predictions composed under \textbf{project-twice} strategy for (a) mean, (b) median, and (c) weighted.}
    \label{fig:2proj}
\end{figure}

\end{document}